\documentclass[final,5p,times,twocolumn]{elsarticle}

\usepackage{amssymb}

 \biboptions{sort&compress}

\usepackage{float}
\usepackage{subfig}
\usepackage[export]{adjustbox}
\usepackage{placeins}
\usepackage{booktabs}
\newcommand{\head}[1]{\textnormal{\textbf{#1}}}
\usepackage{amsmath,bm}
\usepackage{pdflscape}
\usepackage{url}
\usepackage{textgreek}
\usepackage[final]{microtype}
\usepackage{multirow}

\usepackage[usenames]{color}

\usepackage{adjustbox}

\newcommand{\beginsupplement}{%
        \setcounter{table}{0}
        \renewcommand{\thetable}{S\arabic{table}}%
        \setcounter{figure}{0}
        \renewcommand{\thefigure}{S\arabic{figure}}%
        \setcounter{section}{0}
        \renewcommand{\thesection}{S\arabic{section}}
     }

\journal{Powder Technology}

\makeatletter
\def\ps@pprintTitle{%
 \let\@oddhead\@empty
 \let\@evenhead\@empty
 \def\@oddfoot{}%
 \let\@evenfoot\@oddfoot}
\makeatother

\begin{document}

\begin{frontmatter}

\title{Quantitative analysis of thin metal powder layers via transmission X-ray imaging and discrete element simulation: Blade-based spreading approaches}

\author[MIT]{Ryan~W.~Penny\corref{cor1}}
\ead{rpenny@mit.edu}
\author[MIT]{Daniel~Oropeza}
\author[TUM]{Patrick~M.~Praegla}
\author[MIT]{Reimar~Weissbach}
\author[TUM]{Christoph~Meier}
\author[TUM]{Wolfgang~A.~Wall}
\author[MIT]{A.~John~Hart\corref{cor1}}
\ead{ajhart@mit.edu}

\address[MIT]{Department of Mechanical Engineering, Massachusetts Institute of Technology, 77 Massachusetts Avenue, Cambridge, 02139, MA, USA}
\address[TUM]{Institute for Computational Mechanics, Technical University of Munich, Boltzmannstra{\ss}e 15, Garching b. M{\"u}nchen, Germany}

\cortext[cor1]{Corresponding author}

\begin{abstract}
Spreading uniform and dense layers is of paramount importance to creating high-quality components using powder bed additive manufacturing.  Blade-like tools are often employed for spreading powder metal feedstocks, especially in laser powder bed fusion (LPBF) and electron beam melting (EBM), where powders are characterized by a D$_{50}$ of $30$~\textmu m or greater.  Along with variations in boundary conditions introduced by the layer-wise geometry and surface topography of the printed component, stochastic interactions between the spreading tool and powder result in spatial variations of layer quality that are still not well understood. Here, to study powder spreading under conditions representative of PBF AM, we employ a modular, mechanized apparatus to create powder layers from moderately and highly cohesive powders ($15-45$~\textmu m Ti-6Al-4V and $20-63$~\textmu m Al-10Si-Mg, respectively) with a selection of blade-like spreading tools.  Powder layer effective depth is spatially mapped using transmission X-ray imaging, and uniformity is quantified via a statistical approach.  We first compare layer density, or the effective depth of powder layer, and show that blade geometries with a curved profile lead to increased material deposition.  Second, this approach enables quantification of local fluctuations, or layer defect severity.  For example, we observe that the primary benefit of a V-shaped rubber (compliant) blade, as compared to a $45^\circ$ rigid blade, lies in enabling local deflection of the blade edge to eliminate streaking from large particles, while also increasing deposition (layer density). Additionally, we employ a custom DEM simulation to elucidate the opposing roles of particle density and surface energy with a pseudo-material approach, where the balance of inertial and cohesive forces determine macro-scale powder flowability.  For each alloy density, selected to represent Ti-6Al-4V and Al-10Si-Mg, we find via simulations a critical surface energy beyond which layer density is greatly impaired when powder spreading is performed using a blade.

\end{abstract}

\begin{keyword}
additive manufacturing \sep powder \sep spreading \sep X-ray \sep quality \sep discrete element method
\end{keyword}

\end{frontmatter}

\section{Introduction}
\label{sec:Intro}

Powder spreading is integral to powder bed fusion additive manufacturing (PBF AM), wherein layer uniformity and density strongly influence fusion process dynamics, and porosity and mechanical performance of the fabricated component in turn~\cite{Vock2019, Beitz2019, Ziaee2019, Jimenez2019, Korner2016, Brika2020, Wischeropp2019}.  Powder material, particle shape, size distribution, and contamination all influence the innate flowability of powders for this purpose~\cite{Vock2019, Kiani2020, Avrampos2022}; thus, variability is inherent due to the nature of the feedstock.  For example, in laser powder bed fusion (LPBF) blade-like tools are common for spreading powders with typical size distributions of approximately 15-45~\textmu m (D$_{10}$-D$_{90}$)~\cite{Mindt2016, Vock2019, Brandt2017, Sutton2016}.  Here, the objective of effective spreading is a balance between powder flowability, which decreases in fine powders through increased influence of van der Waals forces~\cite{Yablokova2015RheologicalImplants, Chen2017, Meier2019CriticalManufacturing, Meier2019ModelingSimulations}, and layer thickness. Typical layer thickness values are nominally \textgreater 25~\textmu m, but effectively larger by a factor of $\approx 2$ due to particle consolidation upon melting~\cite{Mindt2016})~\cite{Vock2019, Brandt2017, Sutton2016}.  Larger powders are used for electron beam melting (e.g., 45-106~\textmu m) to prevent electrical charges from disturbing particles~\cite{Gong2014, Korner2016}, whereas finer size distributions of metal alloys and ceramics often find application in binder jet AM to reduce sintering time and give higher final part density~\cite{Budding2013, Bai2017, Ziaee2019}.  In all PBF processes, small layer thickness is desirable for high resolution processing; however, overly thin layers results in streaks from comparatively large particles~\cite{Spierings2011, Nan2018, Mussatto2021} and particle bridging that causes sub-surface pockets of low material deposition~\cite{Chen2017}.  Layer non-uniformity may further arise from disturbance from underlying component topography, mechanical inaccuracy and insufficient rigidity of the spreading apparatus, or damage to the spreading tool~\cite{Kleszczynski2012,Hendriks2019, Dana2019}.  Accordingly, the numerous spreading strategies reviewed below aim to maximize layer uniformity for the subsequent fusion step, and include the use of blades of prescribed geometry and material, rollers, hopper-based material deposition, and post-deposition compaction.  

Experimental approaches highlight the basic phenomena and parameters of interest to understanding powder bed density in PBF AM.  To determine general trends in layer density, a common technique is to encapsulate a powder specimen in-situ by fusing a surrounding shell~\cite{Jacob2016, Meyer2017, Wischeropp2019, Chen2019, Brika2020}.  Layer density, as averaged over the gauge volume, is readily assessed through determining the container volume and the mass of powder within.  Using this approach, Brika and coworkers~\cite{Brika2020} demonstrate that spherical powders pack more densely than modestly oblate powders, and demonstrate that size distributions with a high proportion of fine particles decrease layer density in blade-spread 316~stainless steel.  Mussatto~\cite{Mussatto2021}, using a similar approach with Ti-6Al-4V, demonstrates that rheological characteristics of irregularly shaped water atomized powders are unsuitable for PFB AM; this is also discussed in~\cite{Boley2016MetalExperiment, Tan2017AnProcess}. Powder with a large number of fine (\textless~25~\textmu m) particles is also shown to feature higher flow resistance due to particle clusters locking, and increased sensitivity to flow rate.  Thus, more dense layers can be created with this powder, as compared to a slightly coarser powder, but only by limiting spreading speed of the blade below 80 mm/s.  This observation contrasts the aforementioned observations on particle size by Brika, highlighting the additional importance of material and effective surface energy when spreading powders of otherwise similar size distribution and sphericity.

Optical imaging of powder layers permits investigation of fine-scale, surface-level deposition irregularities \cite{Snow2019, Liu2019, Chen2020, Zhang2016, Hendriks2019}.  Snow and coworkers, for example, use an ANOVA approach to generate regressions predictive of layer parameters of Al-10Si-Mg as a function of layer thickness, recoating speed, recoating blade material (also see~\cite{Dana2019}), and powder flowability~\cite{Snow2019}.  While minimal description of the blade geometry is provided, the authors report that percent coverage, optically measured via camera, nonlinearly depends on powder flowability as well as a cross term comprising powder flowability and blade material.  The authors conclude that compliant (silicone) blades may improve percent coverage when using powders of poor flowability, and recommend a rigid implement for freely-flowing powders.  Phuc and Seita \cite{TanPhuc2019AManufacturing} demonstrate high spatial resolution imaging and topography recovery using a contact image sensor, and first demonstrated this in the context of assessing blade damage via powder layer imaging.  This technique is extended to the comparison of rigid and compliant blade spreading of new and recycled 316~SS particles by Le et al.~\cite{Le2021}.  The latter work describes how moderate blade velocities ($10$ to $80$~mm/s) provide the most dense and uniform layers, where insufficient spreading speeds (e.g., $0.2$~mm/s) are associated with severe particle agglomeration.  However, these trends are less severe when using a compliant blade, in part because large particles that are prone to jamming against a rigid blade are able to pass under the compliant blade by deflecting it.  We do note, however, that these methods sense surface topology, which is not directly analogous to resolving spatial variation in material deposition, as the underlying packing fraction varies due to particle size segregation~\cite{Ali2018OnProcesses, TanPhuc2019AManufacturing}, particle bridging~\cite{Chen2017}, and powder compaction~\cite{Penny2021}.

To address the limitations of optical interrogation of powder layer uniformity, X-ray methods are also used \cite{Muniz2018, Escano2018, DuPlessis2018, Brika2020, Heim2016, Schmidt2020, Penny2021}.  One approach is to stabilize powder specimens for ex-situ CT (computed tomography).  This is perhaps best demonstrated by Ali and colleagues~\cite{Ali2018OnProcesses}, who reveal size segregation, or preferential deposition of fine powder particles, in binder-fused powder specimens from a bed of Hastelloy X powder.  Beitz and colleagues employ a similar technique to measure surface roughness of nylon (PA12) layers as related to blade geometry, and observe that increasing the surface area of contact between the blade and powder layer (e.g., a wide, flat blade) minimizes surface roughness.  Alternatively, Escano uses a synchrotron X-ray source to image spreading of a $5$~mm wide powder layer in a side on fashion; it is observed that powder clusters spontaneously form and disintegrate during spreading, and that fine ($D_{50} = 23$~\textmu m v. $D_{50} = 67$~\textmu m), albeit still readily flowable, 316~SS powders result in smoother layers ($R_a = 20$~\textmu m, $R_a = 37$~\textmu m, respectively)~\cite{Escano2018}.  Our recent work proves the utility of X-ray transmission imaging, as described below, to map the effective depth of powder layers under precisely controlled boundary conditions at resolution and scale germane to PBF AM~\cite{Penny2021}.  Across a selection of Ti-6Al-4V, 316~SS, and Al-10Si-Mg powders, the average packing density of single layers is shown to be well-predicted by the angle of repose of the powder and the normalized layer thickness (nominal powder layer thickness divided by powder $D_{50}$).  Variance of packing density within a layer is similarly understood to depend on powder attributes and layer thickness, with large powders resulting in more unpredictable deposition.  Moreover, variation is a function of size scale and orientation, i.e., variance arising from some deposition irregularities such as streaks from large particles is spatially correlated.  Finally, spreading of a new powder layer upon an underlying powder layer introduces additional effects, where spreading forces may induce settling of previously deposited particles into a more dense configuration, thereby increasing the volume of powder deposited in the second layer.  Follow-on work led by Oropeza further applies this experimental method to roller-based spreading of ceramic (alumina) powders, concluding that roller counter-rotation improves uniformity when using finer, more cohesive powder size distributions but impairs layer uniformity for a larger and less cohesive feedstock~\cite{Oropeza2022}.

Simulations have likewise proven powerful in understanding the co-dependence of specific powder traits and spreading strategy on layer parameters.  For example, Haeri~\cite{Haeri2017, Haeri2017_2} studies the effects of blade geometry and powder particle aspect ratio, concluding that a blade with an optimized corner geometry may be superior to a roller-based strategy for realistic, aspherical particle morphologies.  Simulations by Parteli and Chen, studying PA-12 Nylon and 316~SS, respectively, show similar quadratic trends towards decreasing layer thickness and increasing surface roughness with increasing translation speed in roller-spread layers~\cite{Parteli2016, Chen2020}.  Earlier work by Chen, using a $90^\circ$ rigid blade finds that the particle size distribution strongly influences layer density, and that density may be further improved by utilizing bi-modal size distributions~\cite{Chen2019}.  Finally, we note the work of Wang and coworkers who model a $25.9-52.7$~\textmu m nickle-based alloy feedstock, and determine that accurately capturing sliding friction and and van der Waals forces are critical to high-fidelity simulations~\cite{Wang2021}.  Average deposition, randomness (variation), and recoating forces are assessed for a variety of implements, with all blade-like implements showing comparable performance.  However, a cylindrical blade geometry is associated with increased deposition, and high transient forces from particle compression.  Wang et al.~also highlight the complex influence of powder adhesion, noting that increased particle adhesion mitigates particle size segregation by keeping fine particles suspended in the flow~\cite{Wang2020}.  Similarly, our prior work led by Meier~\cite{Meier2019CriticalManufacturing} studies cross-interactions of powder, substrate, blade, and layer thicknesses in blade spreading of Ti-6Al-4V.  Critically, cohesive forces for fine powders (e.g., with a median particle diameter at or below $17$~\textmu m) are observed to be two orders of magnitude greater than gravitational forces, resulting in poor layer quality across nearly all conditions studied.  Additionally, the standard deviation of the packing fraction (as evaluated via binning the packing fraction field on a $100$~\textmu m grid) increases with cohesion, with thinner layers being affected to a larger degree.

\begin{figure*}[htbp]
\begin{center}
	{
	\includegraphics[trim = {1.45in 2.6in 1.5in 2.6in}, clip, scale=1, keepaspectratio=true]{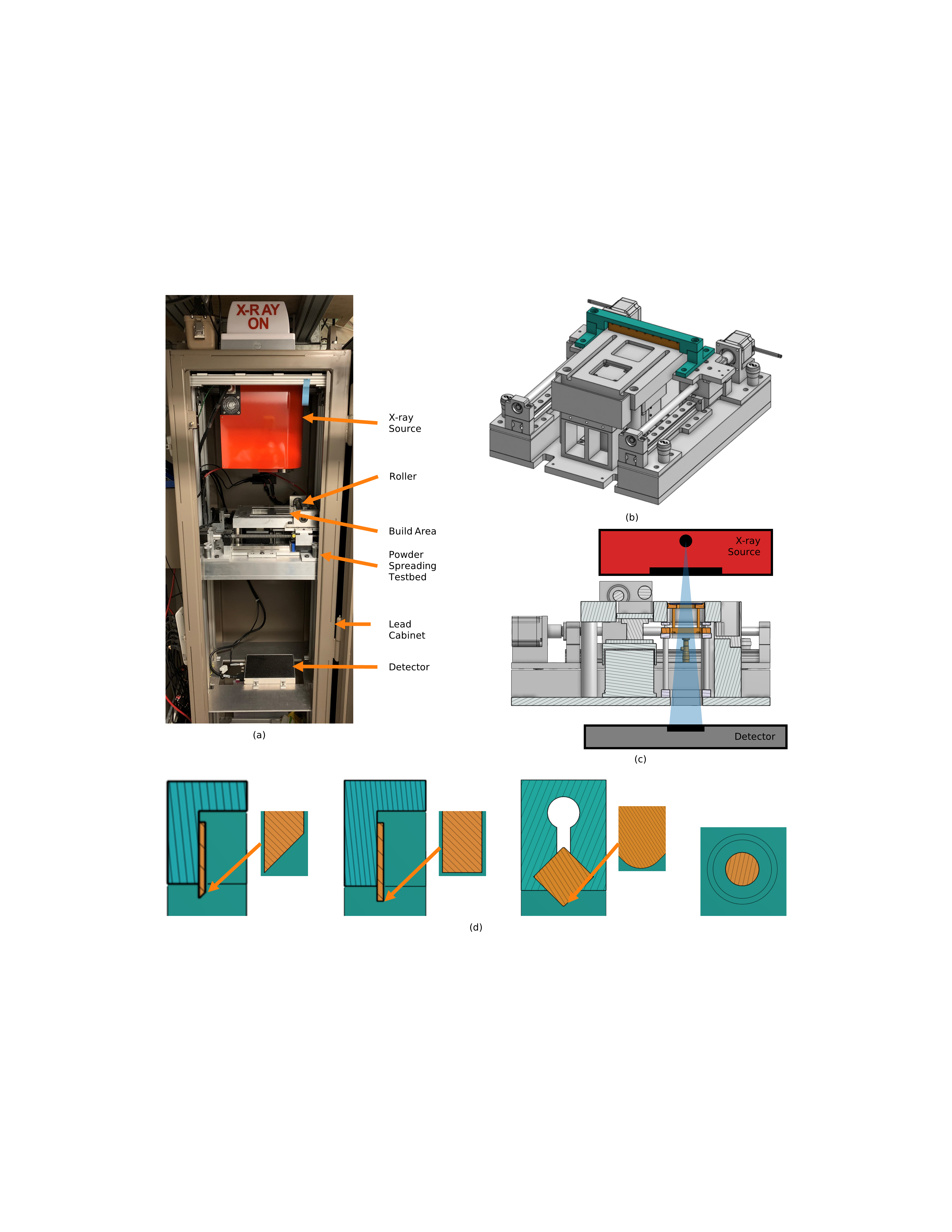}
	}
\end{center}
\vspace{-11pt}
\caption{X-ray powder spreading experiment.  (a) Physical experiment with key components labeled.  (b) Prospective illustration of the powder spreading testbed.  (c) Schematic illustration of X-ray beam path though the X-ray compatible build platform of the spreading testbed.  (d) Cross-sectional view of spreading implements.  From left to right: $45^\circ$, $90^\circ$, V, and cylindrical blades.}
\label{fig:Intro}
\end{figure*}

To further understanding of this complex intersection of powder attributes and spreading mechanics, we employ mapping of mechanically-spread powder layers via X-ray imaging as summarized in Fig.~\ref{fig:Intro}. Powder layers are formed using tools of varying geometry and compliance, each attached to a mechanized spreading apparatus.  Transmission X-ray imaging is then used to spatially map the effective depth of the model powder layers.  For example, Fig.~\ref{fig:IntroLayers}a, shows how a layer of $15-45$~\textmu m Ti-6Al-4V spread with a $45^\circ$ rigid implement shows nonuniformity arising from both particle streaking and blade vibration.  Upon this basis, layer uniformity is statistically quantified, including ascribing layer irregularities to length scale and orientation via power spectral density analysis.  Cohesion is also shown to strongly influence these statistical attributes, both experimentally via study of a high-cohesion Al-10Si-Mg powder and through purpose-built DEM spreading simulations that isolate this parameter for direct study.

\section{Methods}
\label{sec:Methods}

\subsection{Mechanized Powder Spreading}

Powder layers are created using a mechanized spreading apparatus illustrated in Fig.~\ref{fig:Intro}b. The design, mechanical details, and qualification are described in Oropeza et al.~\cite{Oropeza2021}.  To summarize, the spreading tool traverses across a powder supply well and then a simulated build area, driven by two stepper motors coupled to a leadscrew assembly. The spreading speed is typically $50$~mm/s for the experiments described herein. The apparatus further comprises an automated powder supply that uses a motorized piston to meter powder for spreading.  For spreading each layer, approximately four times the theoretical volume of powder necessary to cover the build platform to a depth of $100$~\textmu m is used, as to account for filling gaps in the apparatus (e.g., between the simulated build area and surrounding structure), as well as powder spreading out along the blade as the powder supply is pushed forward.

Figure~\ref{fig:Intro}d illustrates the four recoating tools that may be attached to the testbed.  Specifically, two stainless steel blades are studied: one with a $45^\circ$ knife edge and one with a $90^\circ$, $2$~mm thick edge geometry.  The third tool is a V-shaped compliant blade, wherein a flexure fixture is used to clamp a nitrile (Buna-N) 1/2 in. square O-ring stock at a $45^\circ$ angle (McMaster-Carr, 9700K16).  The corner radius of the o-ring was measured to be $224$~\textmu m, by confocal microscopy (Keyence VK-X1050). To simulate the V blade, using the framework described in Section~\ref{sec:MethodsDEM}, we estimate the elastic modulus from the specified durometer (70A) as $5.5$~MPa using the relationship from Gent~\cite{Gent1958}.  The fourth tool is a steel cylinder ($20$~mm diameter). The rotation axis of the cylinder is fixed, resulting in blade-like spreading. Powder spreading using a rotating roller is studied in a parallel publication.

To enable use of the X-ray microscopy technique described below, a thin plate powder platform mechanism shown in Fig.~\ref{fig:Intro}c is used.  This mechanism supports a 2~mm thick, 6061 aluminum plate, the vertical position of which may be adjusted via two micrometer drives to set the layer thickness.  The micrometers are set by positioning the recoating tool at the center of the plate, and using a gap gauge (Starret, 467M) to obtain a 100~\textmu m offset, thereby determining the nominal powder layer thickness.

\subsection{Powders}
\begin{figure}[htbp]
\begin{center}
	{
	\includegraphics[trim = {3in 4.4in 3in 4.4in}, clip, scale=1, keepaspectratio=true]{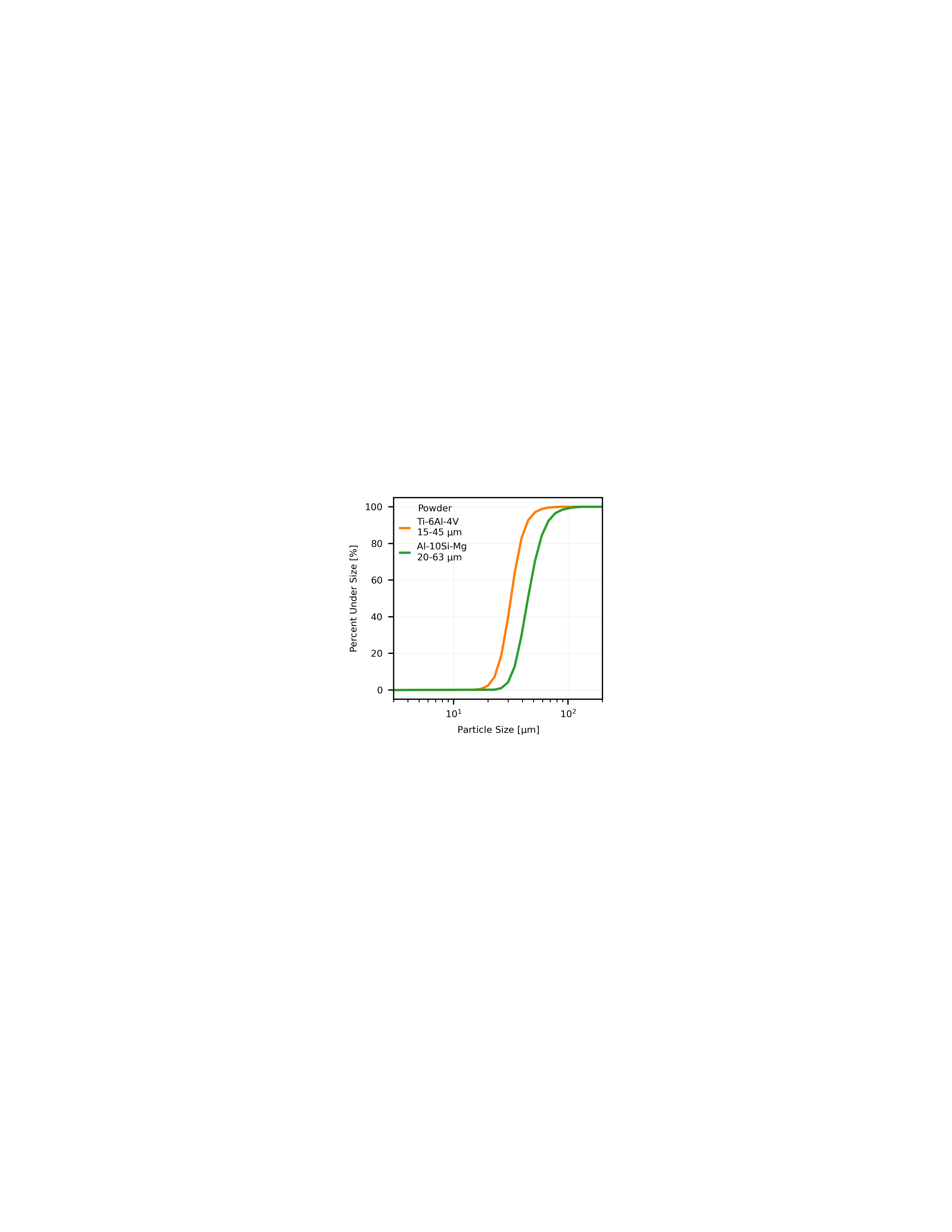}
	}
\end{center}
\vspace{-11pt}
\caption{Cumulative size distributions of the powders used herein, as measured via laser diffraction.}
\label{fig:MethodPowder}
\end{figure}

We study spreading of two powders, as described in Table~\ref{table:Method_Powders} and Figure~\ref{fig:MethodPowder}.  First, we use 15-45~\textmu m Ti-6Al-4V powder as a moderately flowable powder, typical of those employed in metal PBF AM. For this powder we measure an angle of repose (AoR) of $35.6^\circ$ per the method in~\cite{ASTM2013_ReposeAngle} and calculate a Hausner ratio~\cite{Hausner1967} of 1.12 based upon the manufacturer's reported pour and tap densities.  Al-10Si-Mg powder is also studied, with a $20-63$~\textmu m distribution, as to understand the effect of increased powder cohesion.  It features a higher AoR and Hausner ratio ($38.2^\circ$ and 1.44, respectively).

\begin{table*}[htbp]
\centering
\footnotesize
\caption{Tabulated powder properties.}
\label{table:Method_Powders}
\begin{tabular}{@{}*8c@{}}
  
  \toprule[1.5pt]
  \multicolumn{1}{c}{\head{Material}} &
  \multicolumn{1}{c}{\head{Nominal Size [\textmu m]}}&
  \multicolumn{1}{c}{\head{D10 [\textmu m]}} &
  \multicolumn{1}{c}{\head{D50 [\textmu m]}}&
  \multicolumn{1}{c}{\head{D90 [\textmu m]}}&
  \multicolumn{1}{c}{\head{Angle of Repose [$^\circ$]}}&
  \multicolumn{1}{c}{\head{Hausner Ratio}}&
  \multicolumn{1}{c}{\head{Supplier}}\\
  
  \cmidrule{1-8}
 
    Ti-6Al-4V & 15-45 & 23.4 & 31.5 & 43.4 & 35.6 & 1.12 & AP\&C (GE) \\
    Al-10Si-Mg & 20-63 & 32.7 & 44.7 & 64.9 & 38.2 & 1.44 & IMR Metal Powders\\
 
  \bottomrule[1.5pt]
\end{tabular}
\end{table*}

\subsection{X-ray Mapping of Powder Layer Effective Depth}

Powder layer effective depth is mapped using the technique outlined in our prior work, based upon measurement of the X-ray transmission through the layer and the supporting substrate~\cite{Penny2021}.  Figure~\ref{fig:Intro}a shows the apparatus, comprising a Hamamatsu L12161-07 X-ray source (set to $50$~keV, $200$~\textmu A), a Varex 1207NDT flat-panel detector ($18,000$~ms integration time), and the mechanized spreading apparatus, all located within a lead cabinet (Hopewell Designs).  Experiments are performed by first recording an X-ray image of the empty (bare) build region, then spreading the layer, and then recording a second X-ray image of the same area.  Transmission of the powder is then simply determined by taking the ratio of the two images, after interpolation and subtraction of periodically-recorded dark current.  To ensure adequate fidelity and suppress shot noise, each X-ray image comprises a sum of 52 individual frames when imaging Ti-6AL-4V layers, and 520 for Al-10Si-Mg.  This provides an adequate signal-to-noise ratio to ensure $1$~\textmu m uncertainty in the effective depth of powder sensed by each pixel.

Effective depth is graphically defined in Fig.~\ref{fig:IntroLayers}e, and represents the the local thickness of material if the powder were fully densified. This value is determined from the transmission measurements using a radiation transport (RT) model, described thoroughly in our prior work~\cite{Penny2021}.  The RT model begins by calculating the emission spectrum of the X-ray source after Birch and Marshall~\cite{Birch1979}.  This is multiplied by spectral attenuation curves calculated for each item in the beam path, including guards on the X-ray source, a powder layer of specified effective depth, build platform, and detector cover.  These attenuation spectra are calculated using Lambert's Law~\cite{Lambert1760}, using the combination of item elemental composition and thickness.  Attenuation coefficients are determined using elemental data from NIST~\cite{Saloman1988X-ray92} and a rule of mixtures approach~\cite{Saloman1988X-ray92, Dyson1990}.  Finally, the X-ray detection process is incorporated, including representation of X-ray to visible light conversion~\cite{Holl1988} (via cesium iodide scintillator) and detector gain.  With these model components, observed transmission can be densely calculated as a function of powder layer effective depth.  Inversion of this function then permits determination of effective depth from the transmission images collected per the procedure above.

\subsection{Power Spectral Density Analysis}
\label{sec:MethodsPSD}

Uniformity of powder layers can be analyzed from X-ray transmission data via a power spectral density (PSD) frequency-domain approach, which ascribes variance to length scale and orientation~\cite{Penny2021}.  The PSD is computed in three steps.  First, the average is subtracted from the effective depth map, and, second, the two-dimensional Fourier transform is computed.  Finally, the resulting complex amplitudes are squared, resulting in a real-valued representation of power, or equivalently variance, in two-dimensional frequency (inverse dimension) space.  This process maps variance associated with a range of inverse dimensions to an annulus about the origin of the 2D plot, with variance from large, low inverse dimensions appearing closer to the origin and variance from high inverse dimensions being mapped at a larger radius close to the perimeter.  Thus, integrating over an annulus associated with a size scale, taken here to be the frequency bins of the original Fourier transform, yields the total variance associated with effective depth fluctuations between the spatial dimensions associated with the bin edges.  Orientation is also preserved. Thus, restricting the integration to a sector (i.e., to within $\pm 15^\circ$ of the spreading and transverse directions) also allows for variance to be understood as the powder layers are crossed in a specific direction.

\subsection{DEM Modeling}
\label{sec:MethodsDEM}

Discrete element method (DEM) simulations are performed to complement experimental understanding, and to isolate the influence of specific powder parameters beyond the scope of experiments.  Our DEM implementation, described in detail in~\cite{Meier2019ModelingSimulations,Meier2019CriticalManufacturing, Meier2021GAMM, Penny2021}, is based upon the research code BACI~\cite{Baci}.  Spherical powder shapes are assumed, as to simplify particle-particle and particle-boundary (e.g., spreading tool) interactions.  As such, interactions consider normal forces, described via a spring-dashpot model, adhesion from van der Waals attraction, and tangential (frictional) forces, captured via a spring-dashpot model coupled to the normal force with Coulomb's law.  Rolling resistance is also considered in the momentum balance of each particle. The model can also include deformation of the contacting surfaces, allowing the simulation of compliant blades. This is achieved through two-way-coupling of DEM particles and a finite element model of the spreading tool, meaning that forces from the particle-structure interaction are applied to the DEM particles and the tool, and the associated governing equations are solved simultaneously.

The full range of experimental spreading implements is reproduced in simulation (see Fig.~\ref{fig:Simulation}).  Simulations of the  $45^\circ$, $90^\circ$, and V blades spread $21,000$ particles in a nominal thickness of $100$~\textmu m over a domain $1$~mm wide by $7$~mm long.  Periodic boundary conditions are applied along the narrow dimension, such that particles exiting the domain on one side re-enter from the other.  Due to the comparatively large dimension of the cylindrical blade, it is scaled down from a $10$~mm radius in experiment to $5$~mm radius in simulation.  Even with this scaling, simulating a $12$~mm long domain with $42,000$ particles is necessary to accommodate this tool.  These dimensions were determined based on sensitivity studies for the representative volume with each setup.

This DEM framework additionally empowers investigation of how powder flowability and surface energy affect the spreading performance of different blade geometries using a pseudo-material approach.  Our baseline simulation case is modeled on the physical $15-45$~\textmu m Ti-6Al-4V powder studied herein.  Prior work with this powder, both in reproducing scaled-down static angle of repose experiments and manual powder spreading layer densities in~\cite{Penny2021}, suggests a surface energy in the range of $0.04$ to $0.08$~mJ/m$^2$.  Accordingly, and in view of interest in spreading more cohesive powders, we perform simulations where the powder density is held fixed (at a value of $4.43$~g/cm$^3$ corresponding to Ti-6Al-4V) and the surface energy is swept from $0.02$ to $1.28$~mJ/m$^2$ in seven steps on approximately logarithmic intervals.  These values are used to calculate the pull-off force, or the amount of force necessary to split two particles in contact, using the Derjaguin-Muller-Toporov (DMT) model~\cite{Derjaguin1975}.  In turn, the computed pull-off forces are used to capture van der Waals forces that generate powder cohesion in our DEM simulation.  For completeness, identical surface energy sweeps are performed with particles with the lower density ($2.67$~g/cm$^3$) of Al-10Si-Mg.  However, to provide the clearest insight to the effects of density and surface energy, the same $15-45$~\textmu m size distribution is used, instead of the slightly larger distribution of the physical Al-10Si-Mg powder used in experiments.
\section{Results}
\label{sec:Results}

\begin{figure*}[ht]
\begin{center}
	{
	\includegraphics[trim = {1.6in 2.68in 1.1in 3.9in}, clip, scale=1, keepaspectratio=true]{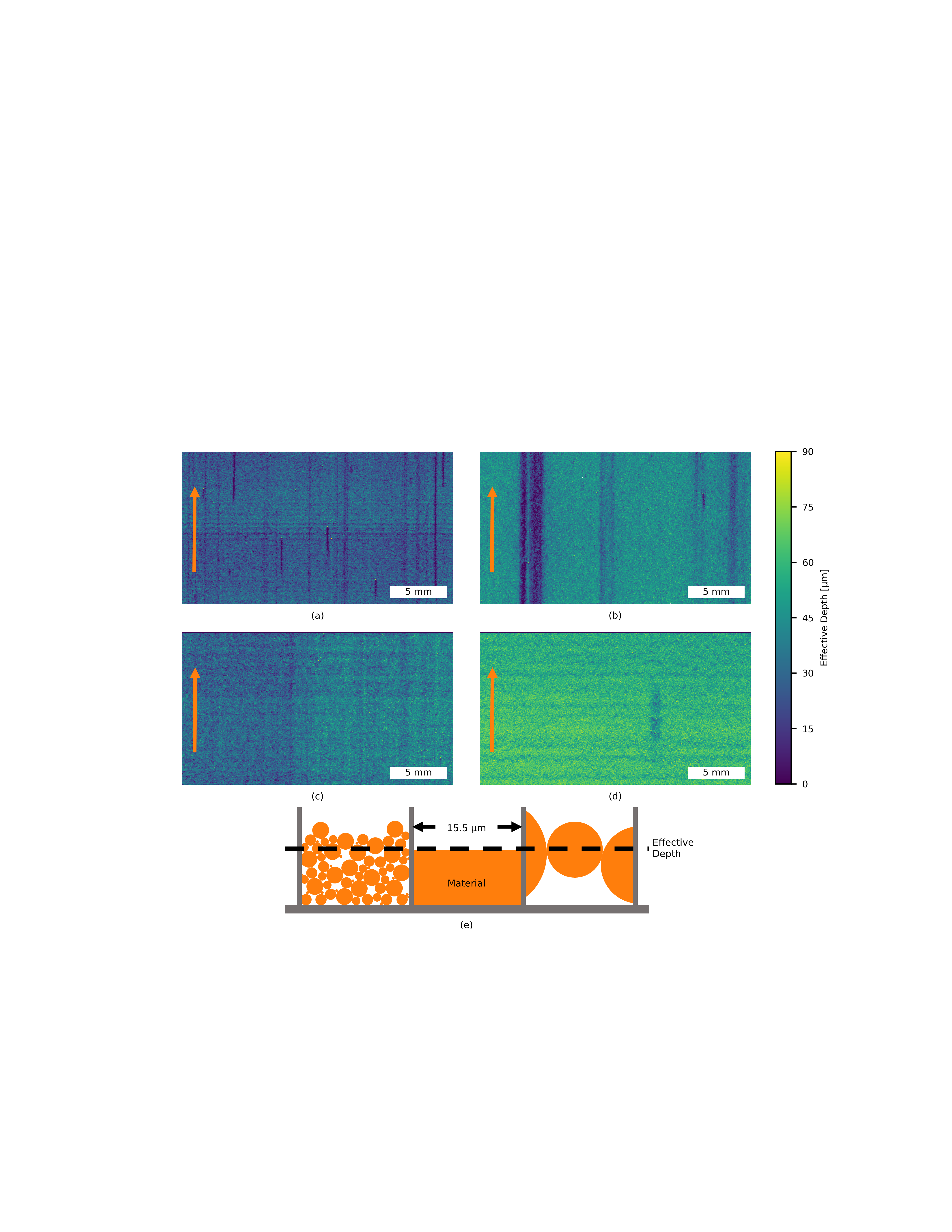}
	}
\end{center}
\vspace{-11pt}
\caption{Eexperimental effective depth images of exemplary $100$~\textmu m thick layers of $15-45$~\textmu m Ti-6Al-4V spread with a: (a) $45^\circ$ blade (24.0~\textmu m effective depth), (b) $90^\circ$ blade (40.0~\textmu m effective depth), (c) V blade (32.8~\textmu m effective depth), or (d) cylindrical blade (58.4~\textmu m effective depth).  Spreading direction is indicated via orange arrows. (e) schematic indicating three material distributions of equivalent effective depth.}
\label{fig:IntroLayers}
\end{figure*}

The primary findings of this study show how the choice of spreading tool influences powder layer packing density and uniformity. Figure~\ref{fig:IntroLayers} illustrates typical 100~\textmu m thick layers of Ti-6Al-4V spread with the range of tools described above, showing maps of the effective layer depth (thickness) created using the transmission model after X-ray imaging.  We begin with the most typical tool used in PBF AM: a rigid blade with a $45^\circ$ knife edge. Layer irregularities are primarily observed as comet-shaped areas of reduced deposition aligned with the spreading direction, where large particles have been unable to pass under the blade and therefore create streak in the layer.  Mechanical vibration of the blade, due to the forces generated by the spreading process in relation to the rigidity of the apparatus, is also evident as slight transverse banding.  Streaking is plainly more severe in the exemplary layer spread using the $90^\circ$ rigid blade.  Specifically, we observe artifacts aligned in the spreading direction, which are practically full-thickness, up to several millimeters wide, and typically extend across the entire length of the area imaged.  However, where powder flow is unimpeded, high effective layer depth is observed.  Thus, the streaking mechanism with the $90^\circ$ rigid blade is slightly different than the $45^\circ$ rigid blade, as the $90^\circ$ geometry allows powder particles to become trapped between the flat surface of the blade and underlying material, simultaneously causing the obvious streaks and deflecting the blade upwards as to increase deposition in the unobstructed areas.  In fact, the forces exerted by the trapped particles are sufficient to damage (scratch) the blade and underlying testbed components.

Conversely, the layer spread using the V blade is nearly streak-free, as even large particles can pass under the edge of the blade by locally deflecting it.  However, increased deposition (e.g., effective depth) is clearly observed in the right half of the imaged area, arising from variation in blade (O-ring stock) dimensions.  Finally, we consider the layer spread with a cylindrical blade; this layer is clearly the densest both here and in the simulations (Fig.~\ref{fig:Simulation}).  Variation is primarily evident as horizontal bands arising from intermittent particle flow, where this stick-slip behavior is made more severe by high normal forces~\cite{Cain2001}, as well as occasional smudge-like defects in the spreading direction.

\subsection{Effect of Tool Geometry on Effective Depth}

Next, we statistically quantify the average and variance of layer effective depth for both experimental and simulated conditions, as shown in Figs.~\ref{fig:ResultsStats} and~\ref{fig:ResultsAl}a.  Beginning with the circles in Fig.~\ref{fig:ResultsStats}, total variance is plotted versus average effective depth, wherein each point represents data from five layers spread with each tool.  Much as expected from the images, powder layers spread with the $45^\circ$ blade are the least dense.  Comparatively, the $90^\circ$ and V blades give more dense layers, albeit with higher variability.  Layers spread with the cylindrical tool have the highest effective depth, due to compaction (downward force) provided by the cylinder geometry, and the lowest variance in proportion to deposition. 

On Fig.~\ref{fig:ResultsStats}, plus signs (+) alternatively indicate layer-to-layer variance, determined here by first computing the variance of effective depth pixel-wise (i.e., the variance of the five effective depth measurements is determined for each sub-region of the powder layer sensed by an individual detector pixel), then averaging the resulting variance values.  Comparing this metric against the former notion of variance provides an understanding of the spatial repeatability of defects.  Specifically, if all layer irregularities were systematic as a function of position this metric would be 0, and if all layer irregularities were random, this metric would be equivalent to the previous notion of layer variance.  In all cases we observe that this metric lies substantially below the total layer variance, indicating that a portion of layer irregularity is a systematic function of position.  However, we specifically note the V and cylindrical blade data, where, on a proportional basis, systematic variation is slightly lower than characteristic of the rigid blade layers.

Additionally, increased particle cohesion is experimentally investigated using the low-flowability (Al-10Si-Mg) powder, and results are shown in Fig.~\ref{fig:ResultsAl}a.  These data indicate much higher variance, approximately $500$~\textmu m$^2$, in layers of this low flowability powder as compared to the Ti-6Al-4V layers, yet otherwise agree, showing slightly higher average, and more variable, deposition with the V blade as compared to the rigid $45^\circ$ blade.

\begin{figure*}[hbt]
\begin{center}
	{
	\includegraphics[trim = {1.6in 4.3in 1.6in 4.3in}, clip, scale=1, keepaspectratio=true]{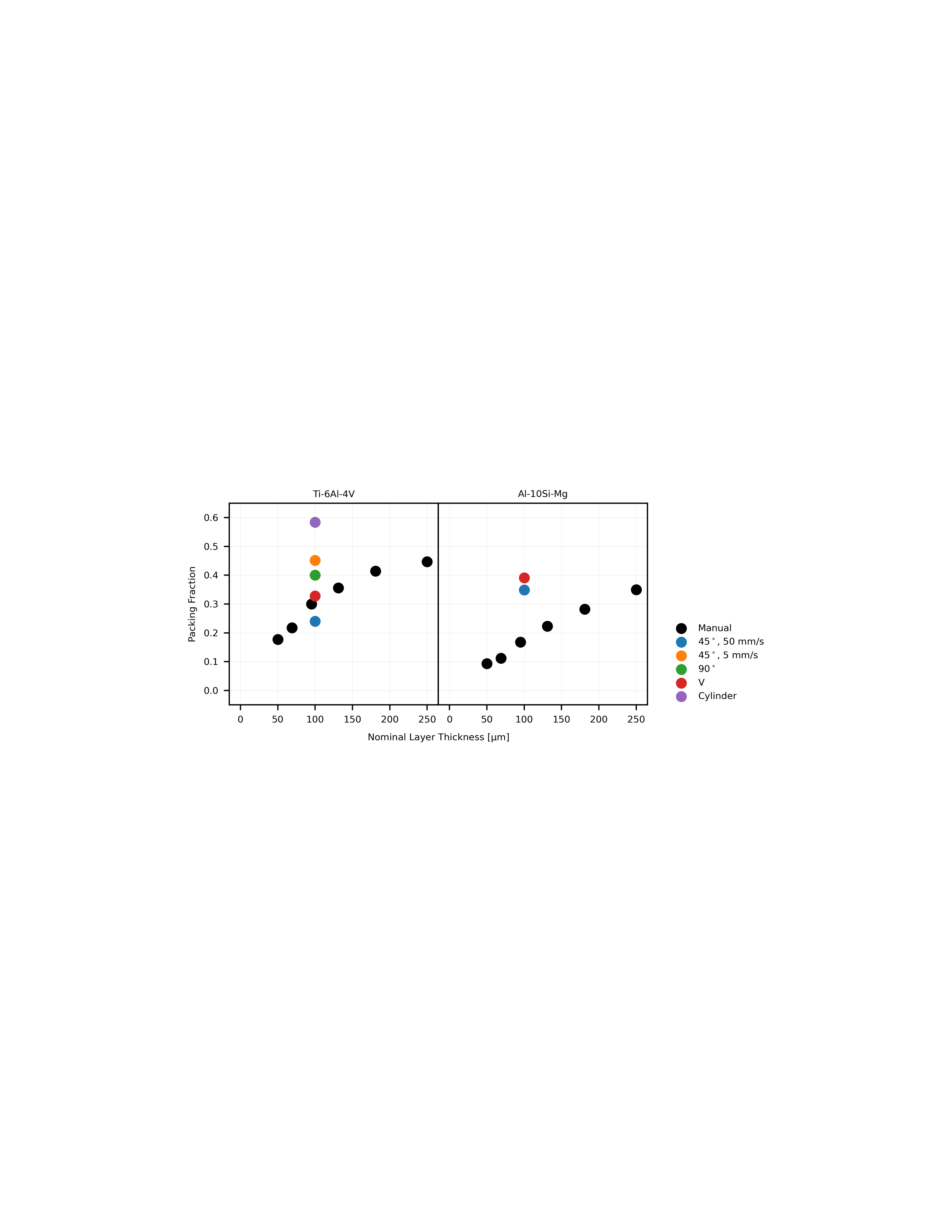}
	}
\end{center}
\vspace{-11pt}
\caption{Comparison of mechanized spreading experiments to manual spreading experiments of the same powders in~\cite{Penny2021}.}
\label{fig:SupplementManual}
\end{figure*}

Casting these results at a packing fraction, or simply dividing the average effective depth by the nominal layer thickness, allows contrasting these results against our prior work with these powders in~\cite{Penny2021}.  To summarize our earlier findings, packing fraction asymptotically approaches a constant value ($0.5$ and $0.59$ for Ti-6Al-4V and Al-10Si-Mg, respectively) with increasing layer thickness, when using a machinist's flat ($1/8$~in. thick, $90^\circ$ edge) to manually spread powder into precision-etched silicon layer templates of a range of nominal layer thicknesses.  Beginning with the Ti-6Al-4V powder in Fig.~\ref{fig:SupplementManual}, the $90^\circ$ implement is the closest in geometry to our earlier work; however, the packing fraction achieved with this implement in this context is well above the manual-spreading trend.  This is due to the powder entrapment and machine deflection that lead to localized yet strongly enhanced powder deposition, as described above.  The high normal forces exerted on the powder by the cylindrical implement likewise lead to an exceptionally high packing fraction of $0.58$.  Notably, this is higher than the asymptotic value from the manual spreading experiments and indicates a substantial degree of powder compaction.  Layers spread with the $45^\circ$ and V blades lie closest to the trend, and the packing fraction of thicker and thinner mechanized spreading is expected to trend similarly to the manual spreading results.  These two mechanized implements appear to increase the deposition of Al-10Si-Mg powder; however, we hypothesize that this is only in part due to the difference in implement geometry.  Our parallel work in roller-based spreading of this powder suggests that, opposite the Ti-6Al-4V powder, higher spreading (shear) forces serve to overcome the high cohesive forces innate to this powder and thereby increase powder deposition.  Thus, a portion of the increased deposition here may be explained by the higher traverse speed ($\approx 5$~mm/s manually versus $50$~mm/s here) exposing the shear-dependent flow characteristics of this powder.  Even with this increase, the packing fractions achieved mechanically lie well below the asymptotic value, suggesting that much room remains to improve the layer density achieved with this powder.

\begin{figure*}[ht]
\begin{center}
	{
	\includegraphics[trim = {1in 3.7in 1in 4.2in}, clip, scale=1, keepaspectratio=true]{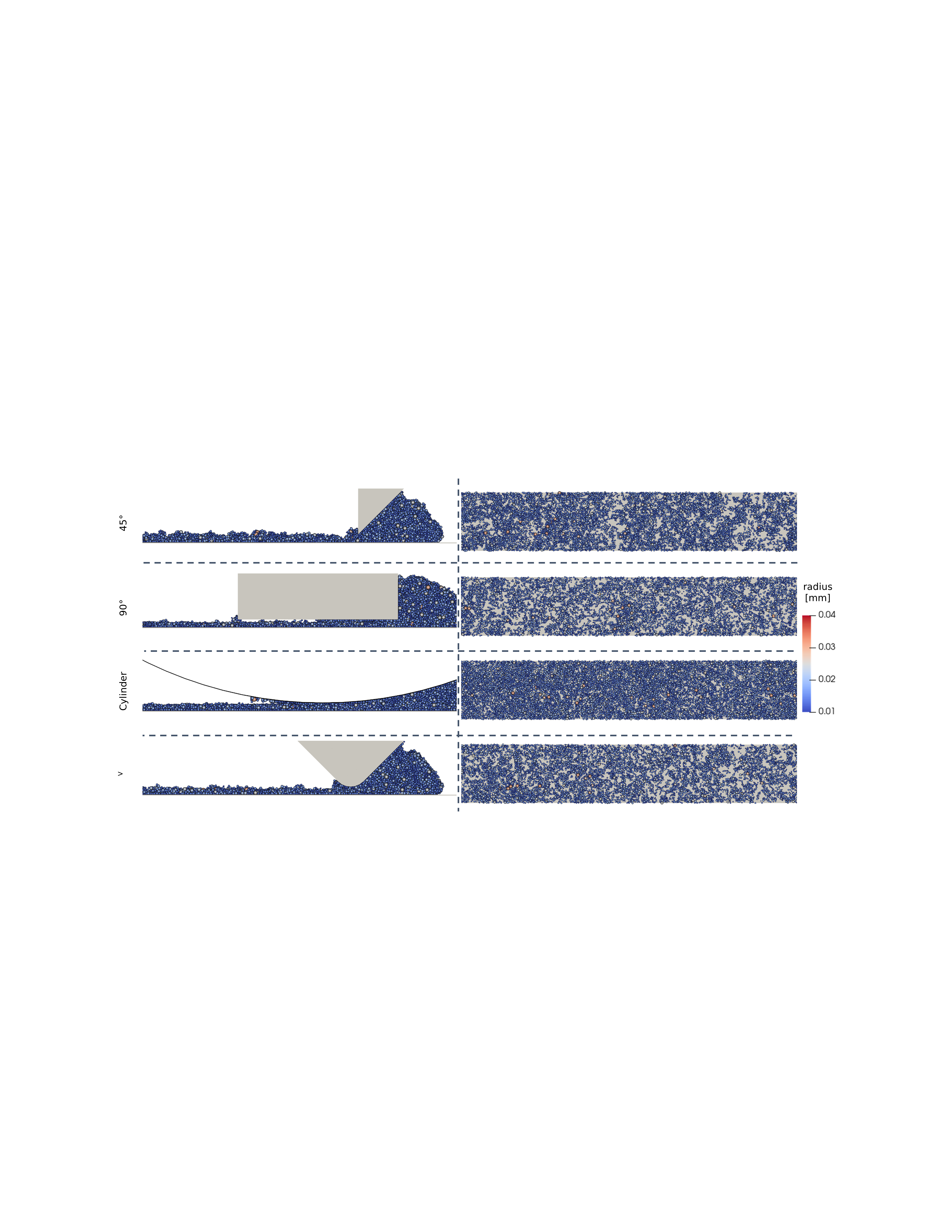}
	}
\end{center}
\vspace{-11pt}
\caption{Exemplary simulation results spreading $100$~\textmu m layers of $15-45$~\textmu m Ti-6Al-4V with a surface energy of $0.32$~mJ/m$^2$ at $50$~mm/s.  Left: Side view of the spreading process, with the tools traversing from left to right.  Right: Top view of the resulting layers showing clear differences in deposition.}
\label{fig:Simulation}
\end{figure*}

Leveraging the DEM simulation, the relationships between spreading implement and specific powder attributes on layer quality may be deduced.  Figure~\ref{fig:ResultsSimsStats}a compares pseudo-materials with a material density representative of Ti-6Al-4V, over the range of deliberately altered surfaces energies motivated in Section~\ref{sec:MethodsDEM} (also see Fig.~\ref{fig:ResultsSimsAdhesion} for simulation results plotted against surface energy).  Comparing the $45^\circ$ and $90^\circ$ blades in these figures, average deposition follows a very similar trend: roughly constant below a surface energy of $\approx 0.3 2$~mJ/m$^2$ and rapidly falling off for more cohesive powders with surface energies greater than this value.  However, the variance of layers fabricated with the latter geometry is lower, coming to a much lower peak in the vicinity of this critical surface energy.  This arises as deposited powder particles continue to interact with the bottom edge of the tool as it passes over, as visible in Fig.~\ref{fig:Simulation}, enhancing the uniformity of the layer.  The magnitude of the variance peak in the cylinder-spread layers falls between the two rigid blade cases, again suggesting a smoothing effect from the larger surface area of the tool in contact with the top surface of the powder layer.  Higher average depth for high and moderate flowablility powders (surface energy $\leq 0.08$~mJ/m$^2$) is also observed, again via simulations, with the cylindrical implement, as the geometry serves to gradually compact powder particles into a more dense configuration as it deposits powder across the simulated build area.  At the high surfaces energies considered here, the cylindrical tool results in deposition similar to the $45^\circ$ and $90^\circ$ blades.  

Finally, we consider two simulations that incorporate the geometry of the V blade used in the experiments; namely, a first set of simulations that captures both the shape and high compliance (via scaled elastic modulus) of the experimental V recoating implement and a second set wherein an implement of identical geometry, but in a low-compliance (i.e., rigid steel) material is simulated.  Results from these two cases are practically indistinguishable, indicating that the average forces of powder particles exerted on the high-compliance (rubber) blade are insufficient to cause any large-scale deflection of the blade edge.  Average deposition in this case most closely resembles the layers created with the cylindrical implement at low surface energy, but roll off more gradually at high surface energy much like the $45^\circ$ and $90^\circ$ blade layers (e.g., at surface energies $\geq 0.32$~mJ/m$^2$ as qualitatively seen in Fig.~\ref{fig:Simulation}).  We conclude that, in the context of industrial practice, the benefit of compliant recoating blades lies in enabling geometries that increase the surface area in contact between the powder and implement, while providing local compliance to avoid large-particle streaking and elevated force transients (as described, e.g., in~\cite{Wang2021}) that may occur with rigid recoating blades of equivalent geometry and/or due to inherent variation in the top surface of the component being printed (e.g., due to residual stress). 

\begin{figure*}[hbt]
\begin{center}
	{
	\includegraphics[trim = {1.6in 3.5in 1.6in 3.2in}, clip, scale=1, keepaspectratio=true]{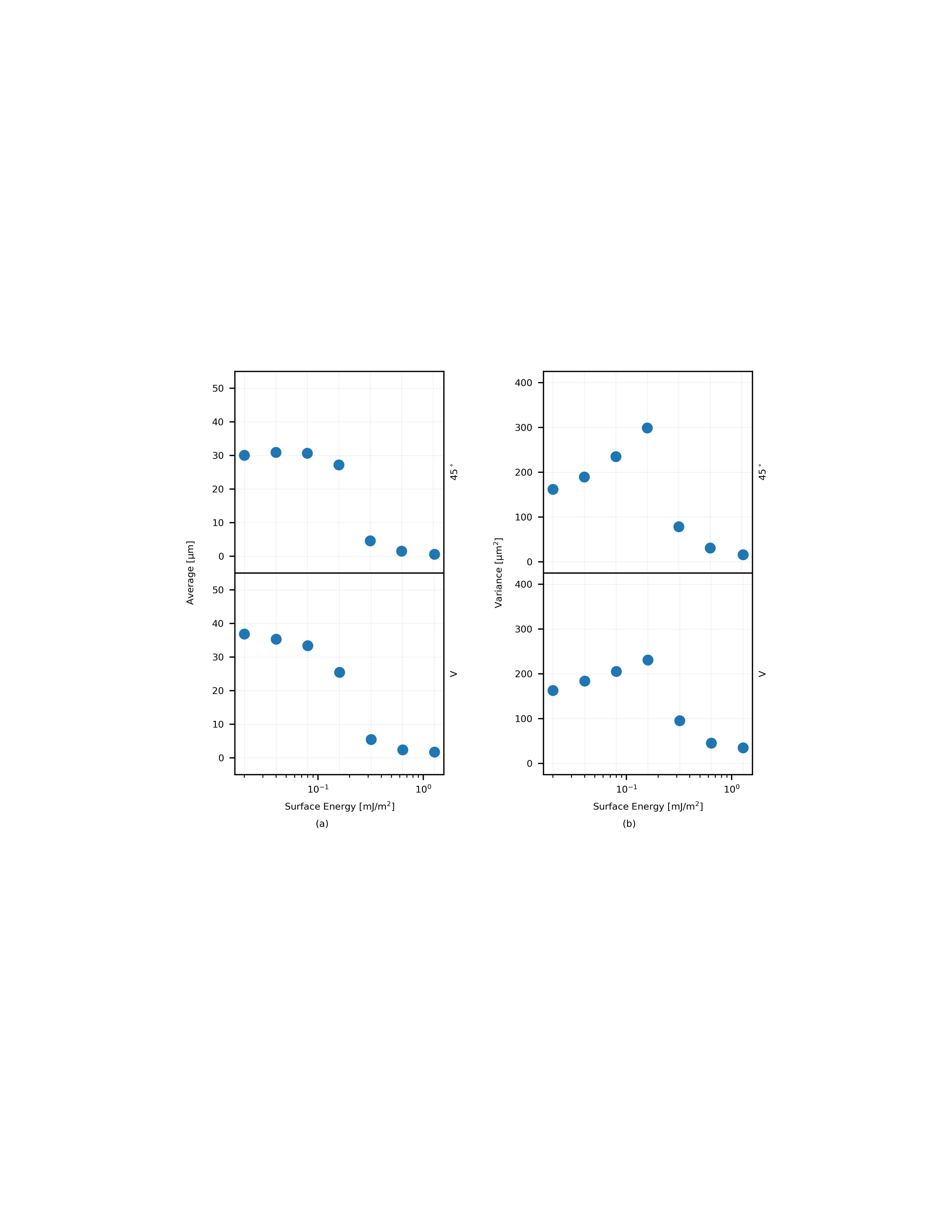}
	}
\end{center}
\vspace{-11pt}
\caption{Simulated layer effective depth statistics versus pseudo-material surface energy, where material density is representative of Al-10Si-Mg.}
\label{fig:SupplementAlStats}
\end{figure*}

Figures~\ref{fig:ResultsSimsStats}b and~\ref{fig:SupplementAlStats} comprise simulation results where particle material density has been lowered to represent Al-10Si-Mg; we note that there is minimal change in the functional trend and magnitude of the average and variance curves.  Rather, the only substantive difference is that the Al-10Si-Mg points lie at slightly different locations (or, more simply, shifted to the left in Fig.~\ref{fig:SupplementAlStats}) for the less-dense powder, indicating that an equal level of cohesive forces will more greatly impact flowability of a powder of less dense material than a more dense one (See~\cite{Meier2019ModelingSimulations} for a more detailed discussion of how these parameters scale).  Further, average layer depth in Al-10Si-Mg spreading experiments closely corresponds to simulations of low-to-moderate cohesion (surface energy of $0.02$ to $0.16$~mJ/m$^2$ in Fig.~\ref{fig:ResultsSimsStats}b). However, the experimentally observed variance is much higher than suggested by the simulations.  We rationalize this difference on two considerations.  First, our prior study conclusively shows that as-spread layers of wider size distributions, all else equal, feature increased variance~\cite{Penny2021} and again the experimental powder here is slightly larger than the simulated version.  Second, the Al-10Si-Mg powder is modestly oblate, which is known to impart higher variance as compared to layers created from spherical powder particles.

\subsubsection{Cumulative Distribution Functions}

\begin{figure}[ht]
\begin{center}
	{
	\includegraphics[trim = {3in 1.6in 3in 1.5in}, clip, scale=1, keepaspectratio=true]{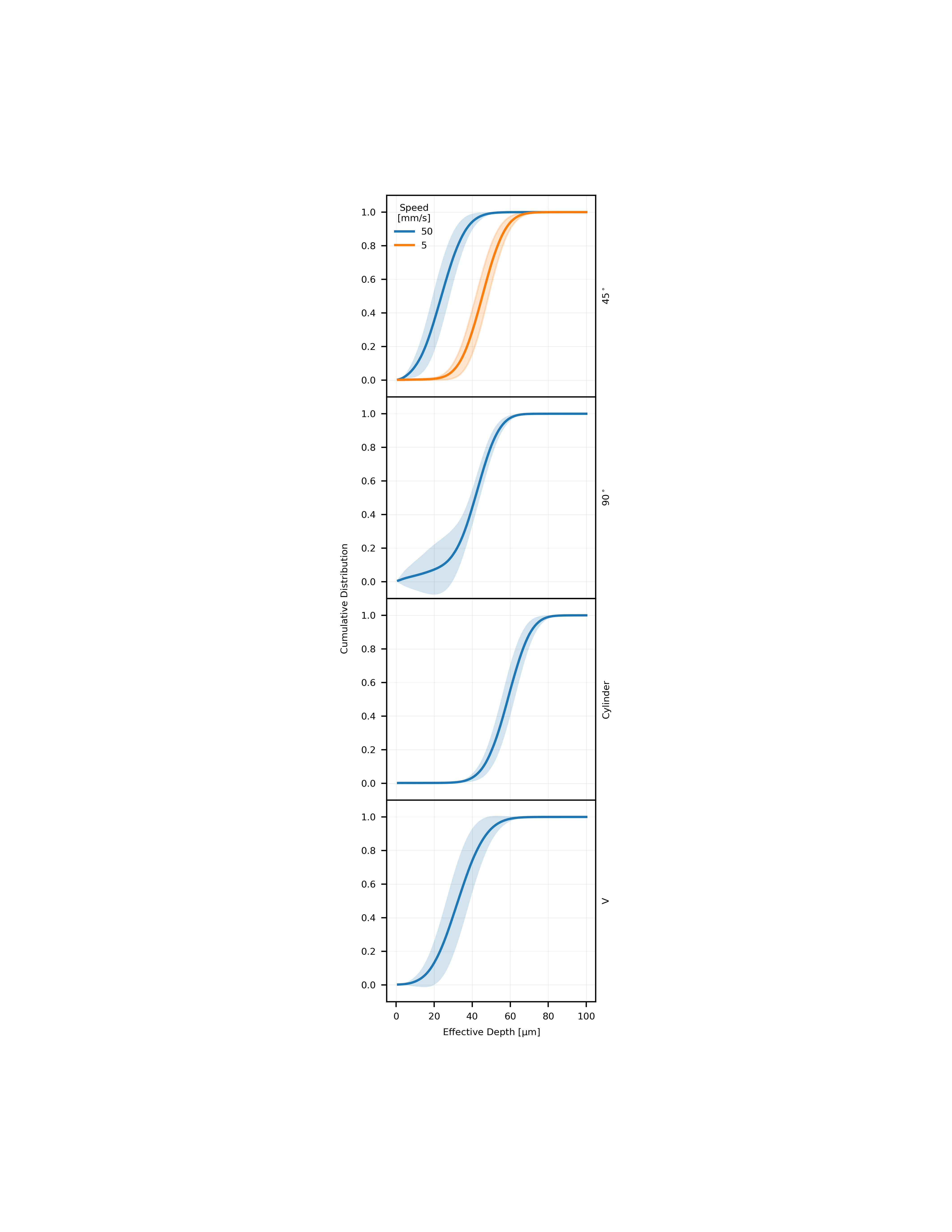}
	}
\end{center}
\vspace{-11pt}
\caption{Effective depth cumulative distribution functions for experimental $15-45$~\textmu m Ti-6Al-4V powder layers created with a selection of spreading implements.  Solid lines are the average of $5$ layers and shaded regions denote $\pm 3 \sigma$ bounds.}
\label{fig:ResultsHists_Expt}
\end{figure}

X-ray transmission measurements enable exploration of the cumulative distribution function (CDF) of layer effective depth, which provides visual representation for density, uniformity, and layer-to-layer variability.  Figure~\ref{fig:ResultsHists_Expt} plots the average cumulative distributions of effective layer depth for each spreading tool/configuration, and the shaded regions additionally represent the $\pm 3$~\textsigma{} bounds at a given effective depth, providing a sense of how consistently the Ti-6Al-4V powder is spread from one layer to the next.  

Layers spread with the $45^\circ$ blade show reasonably consistent layer-to-layer performance. Bare or thinly coated regions are rare due to the small size of the streak artifacts.  Spreading with the $90^\circ$ rigid blade causes a pronounced kink at approximately 30~\textmu m of effective depth, with a comparatively high likelihood of pixels with effective depths spanning $0-30$~\textmu m.  Moreover, the specific number of pixels with a given effective depth is also quite variable in this range.  Yet, deposition is significantly more repeatable in regions of high effective depth occurs.  Streaks are also probable, and while they show some variability in severity, are biased towards full-thickness in depth.  The V and cylindrical blade CDFs are similar to the $45^\circ$ curve.  They indicate slightly more and less variability from layer to layer, respectively, and no substantial bias to a specific effective depth.  Experimental results for single layers created with the higher cohesion Al-10Si-Mg powder (Fig.~\ref{fig:ResultsAl}b) show reduced sensitivity to the choice of spreading tool when comparing the $45^\circ$ and V blades.  Notably, in both cases a higher probability of bare regions is measured relative to the Ti-6Al-4V powder, albeit with a slightly lower probability when using the V blade compared to the $45^\circ$ rigid blade.

\begin{figure*}[ht]
\begin{center}
	{
	\includegraphics[trim = {2in 1.6in 2in 1.7in}, clip, scale=1, keepaspectratio=true]{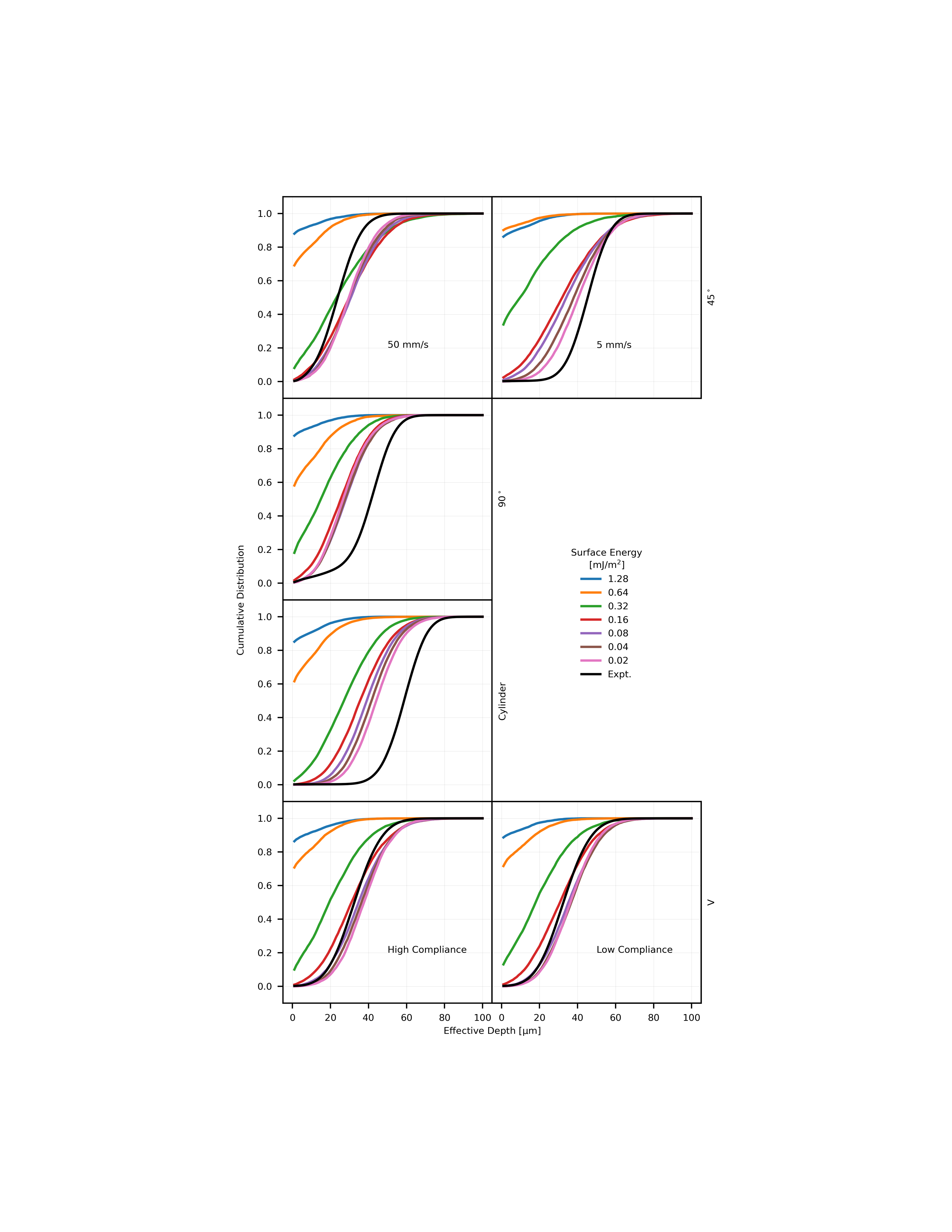}
	}
\end{center}
\vspace{-11pt}
\caption{Effective depth cumulative distribution functions of simulated Ti-6Al-4V layers corresponding to a range of powder surface energies, compared to experiments.  Note: the extra panel using the $45^\circ$ blade implement is at a reduced speed of $5$~mm/s. The bottom panels separate the effects of the geometry of the V blade from its elastic modulus; specifically, the left panel realizes the geometry in a high-compliance (rubber) material versus a low-compliance (steel) material on the right.}
\label{fig:ResultsHists_Sims}
\end{figure*}

Figure~\ref{fig:ResultsHists_Sims} presents simulated CDF data using each tool to spread pseudo-materials with a density matching Ti-6Al-4V and a range of surface energy values (the corresponding experimental data shown in black for comparison).  Starting with the $45^\circ$ panel, we note that the spreading behavior of the comparatively flowable powders ($0.02$ to $0.16$~mJ/m$^2$) are similar, whereas bare regions become increasingly likely beyond this range.  In comparison to the experiment, the curve shapes are quite similar in shape with the greatest difference occurring at the high-deposition (greater than $\approx 30$~\textmu m effective depth) side of the distribution.  The effect of bare and thinly coated regions becomes more obvious in the $90^\circ$ blade panel, specifically in the dog-leg shape of the distributions for more flowable powders at the low-effective-depth end of the curve.  As observed in the experimental data, this distinct curve shape arises from the effect of large particles becoming trapped against the blade, leaving defects of near full-thickness.  Again, the experimental curve is shifted to the right, due to upward deflection of the blade and recoating apparatus by particles trapped under the blade.  The effective depth of simulated layers fabricated with the cylindrical blade show the greatest sensitivity to powder flowability at low surface energies, where the roller shape serves to compact powders of high innate flowability.  However, the increased downward force does not improve deposition of high-surface-energy (approximately $0.64$ mJ/m$^2$ and above) powders, as evidenced by the increased probability of bare regions, with performance resembling the $45^\circ$ and $90^\circ$ blades.  Last, we investigate the layers by the simulated V blade, and again observe that the change in deposition as compared to the $45^\circ$ blade is primarily an effect of geometry and not from gross deflection of the pliable blade edge, in comparing the left and right panels at the bottom of the figure.  Again relative to the $45^\circ$ blade, the V blade features reduced deposition at high surface energies, much like the comparable shape at the implement tip to the roller.  For lower surface energy, more flowable powders, however, deposition increases.  This is driven by a lower probability of moderate-to-low deposition regions e.g., the bottom inflection points of the PSD curves shift right to higher effective depth.

\begin{figure}[hbt]
\begin{center}
	{
	\includegraphics[trim = {3in 3.5in 2.2in 3.4in}, clip, scale=1, keepaspectratio=true]{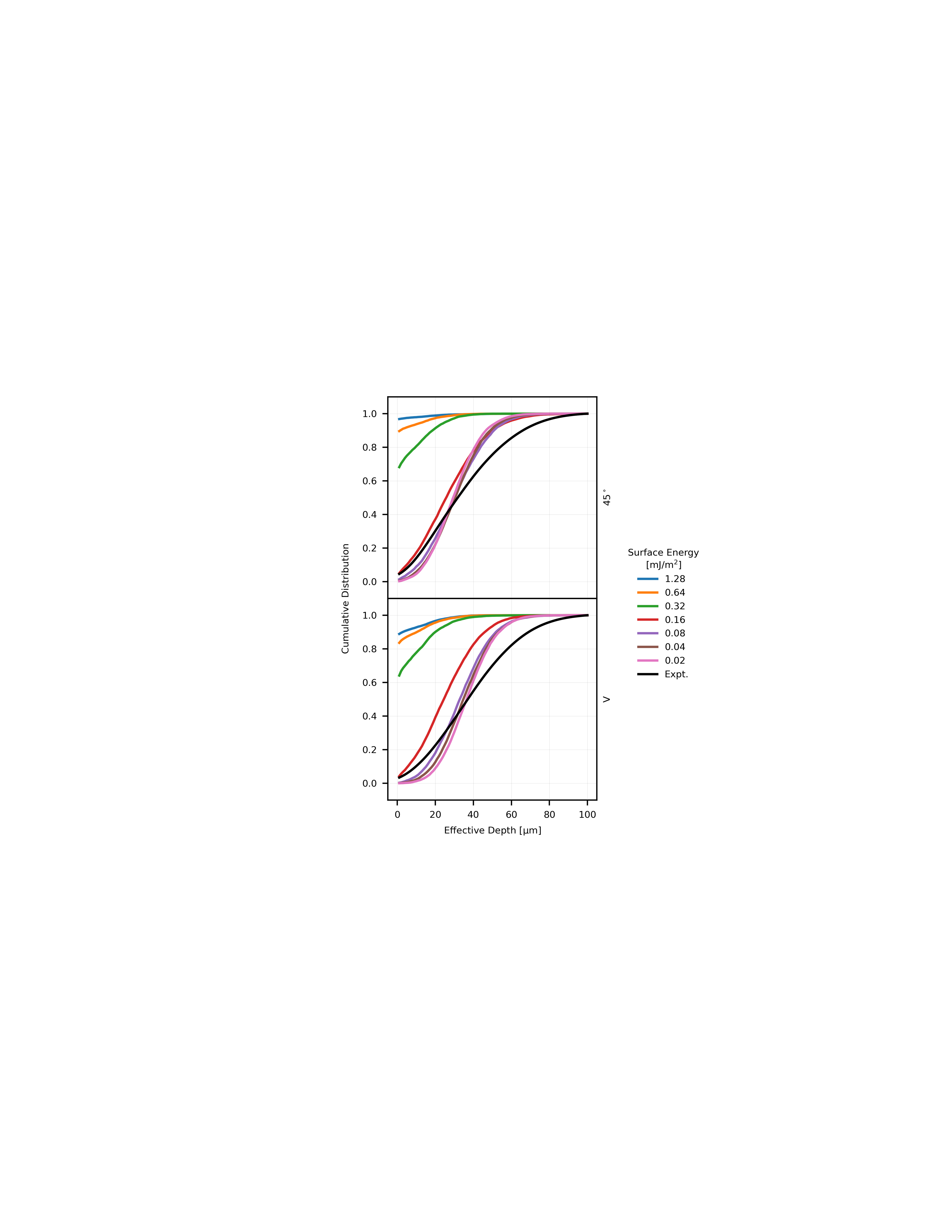}
	}
\end{center}
\vspace{-11pt}
\caption{Cumulative distributions of effective depth for the simulated $45^\circ$ and V blade implements, wherein material density is representative of Al-10Si-Mg.  Pseudo-material surface energies are swept from $0.02$ to $1.28$~mJ/m$^2$.}
\label{fig:SupplementAlHists}
\end{figure}

Simulations for layers of the low density (Al-10Si-Mg) powder spread with the $45^\circ$ and V blades are included in Fig.~\ref{fig:SupplementAlHists} for completeness; these curve families effectively match those in Fig.~\ref{fig:ResultsHists_Sims}, albeit with a slight shift due to the density-driven difference in flowability associated with a given surface energy.

\subsubsection{Power Spectral Density}

\begin{figure*}[ht]
\begin{center}
	{
	\includegraphics[trim = {1in 4.2in 1.44in 4.2in}, clip, scale=1, keepaspectratio=true]{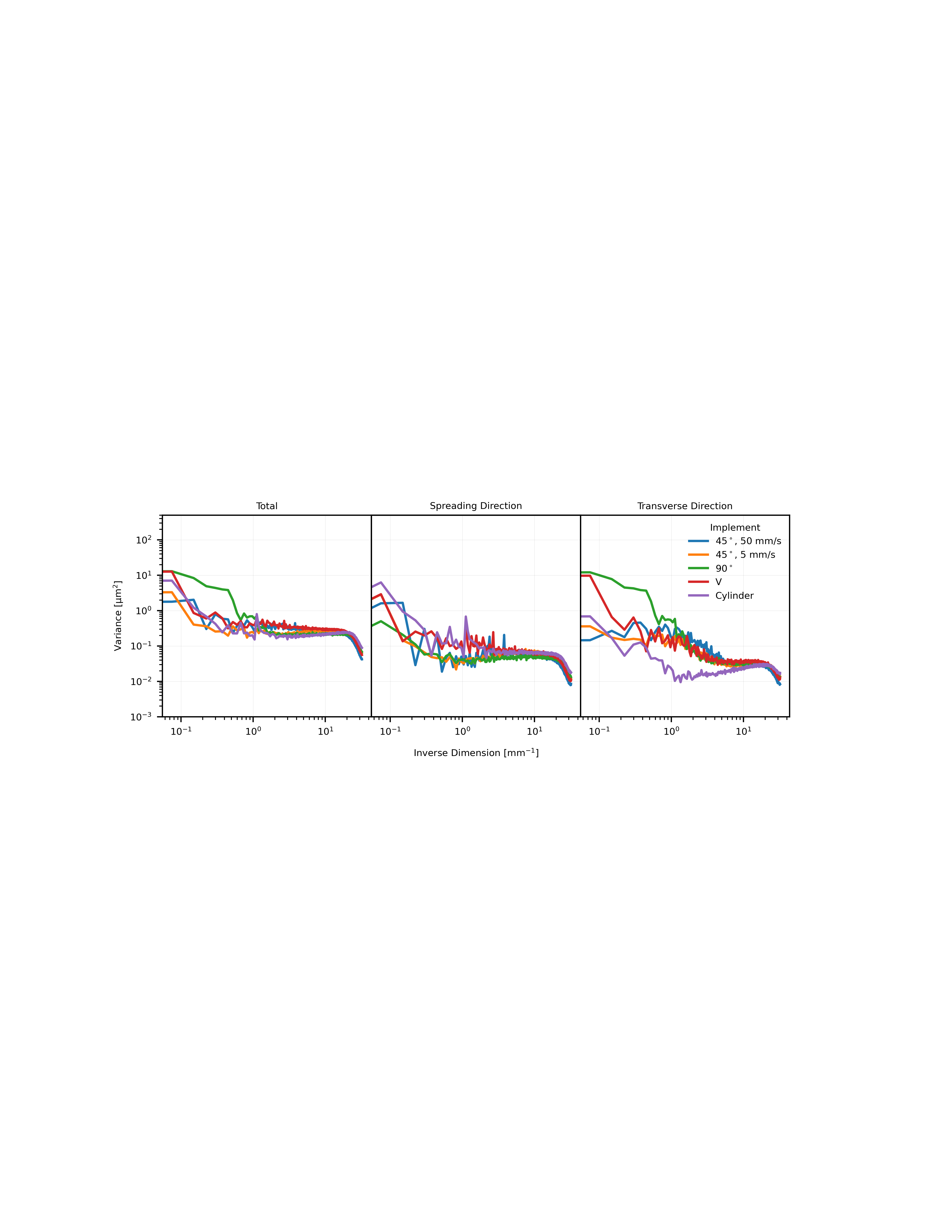}
	}
\end{center}
\vspace{-11pt}
\caption{Average PSD analysis of layer variance as a function of size scale for experimental layers of Ti-6Al-4V, integrating over all directions, within $\pm15^\circ$ of the spreading direction, and within $\pm15^\circ$ of the transverse direction, from left to right.}
\label{fig:ResultsPSD}
\end{figure*}

With the PSD analysis method described in Section~\ref{sec:MethodsPSD}, we further quantify the size, orientation, and severity of defects in experimental layers spread with each tool.  Beginning with the $45^\circ$ rigid blade layers in the left panel of Fig.~\ref{fig:ResultsPSD}, we find that the highest variance is associated with low inverse dimension, is approximately constant at intermediate dimension, and rolls off at high inverse dimension.  Considering the directionally-resolved plots in the center and right, the increase in variance at low inverse dimension (below $0.02$~mm$^{-1}$) is primarily oriented in the spreading direction, and a broad hump occurs in the transverse direction at $0.02$-$5$~mm$^{-1}$), corresponding to the comet-like defects observed in Fig.~\ref{fig:IntroLayers}a.  The larger, more sever streaking visible in layers spread with the $90^\circ$ rigid blade is also easily quantified with this method; namely, the increase in variance compared to the $45^\circ$ blade layers occurs at inverse dimensions below $1$~mm$^{-1}$.  In the spreading direction, variance associated with this implement is is relatively low in most frequency bins below $1$~mm$^{-1}$, as the streaks tend to extend for the entire length of the area imaged and therefore do not cause effective depth to strongly vary in this direction.  Thus, the increased total variance is driven by nonuniformity in the transverse direction, correlating to the wide and deep streaks caused by powder trapped under the blade.  Layers spread with the V blade appear nearly isotropic, aside from increased variance at very low inverse dimensions in the transverse direction due to the form-error of the blade itself.  Last, when the cylinder is used for spreading, variance appears higher in the spreading direction including a clear peak at approximately $1.5$~mm$^{-1}$ due to the intermittent, stick-slip powder flow previously discussed.  In the transverse direction, these layers feature low variance, likely due to the high rigidity and form accuracy of the cylindrical tool.  

\begin{figure}[ht]
\begin{center}
	{
	\includegraphics[trim = {2.3in 2in 3.85in 1.8in}, clip, scale=1, keepaspectratio=true]{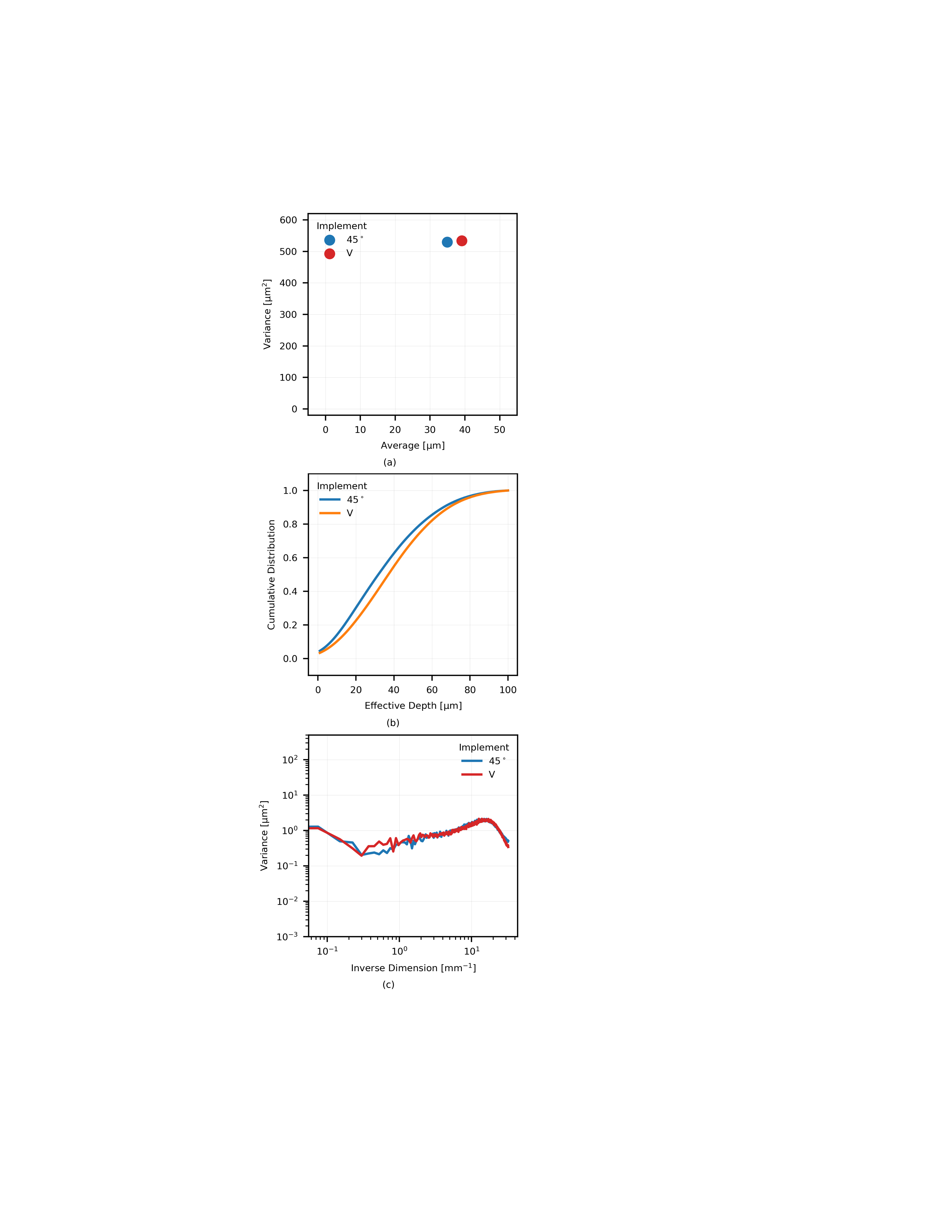}
	}
\end{center}
\vspace{-11pt}
\caption{Analysis of experimental $20-63$~\textmu m Al-10Si-Mg powder layers created with the $45^\circ$ and V blades.  (a) Comparison of layer statistics.  (b) Cumulative distribution functions of layer effective depth.  (c) Power spectral density comparing layer variance as a function of inverse dimension.}
\label{fig:ResultsAl}
\end{figure}

Consistent with the other experimental results, we observe that the low flowability of the Al-10Si-Mg powder masks any effect of the spreading tool on the PSD analysis, as shown in Fig.~\ref{fig:ResultsAl}c.  However, in contrast to the Ti-6Al-4V data, we note a broad peak, centered around $15$~mm$^{-1}$.  In conventional terms, this is $67$~\textmu m or approximately $1.5$ times the mean particle diameter, indicating that powder clusters manifest as a portion of the observed  non-uniformity of Al-10Si-Mg layers.

\begin{figure}[ht]
\begin{center}
	{
	\includegraphics[trim = {3in 4.4in 3in 4.4in}, clip, scale=1, keepaspectratio=true]{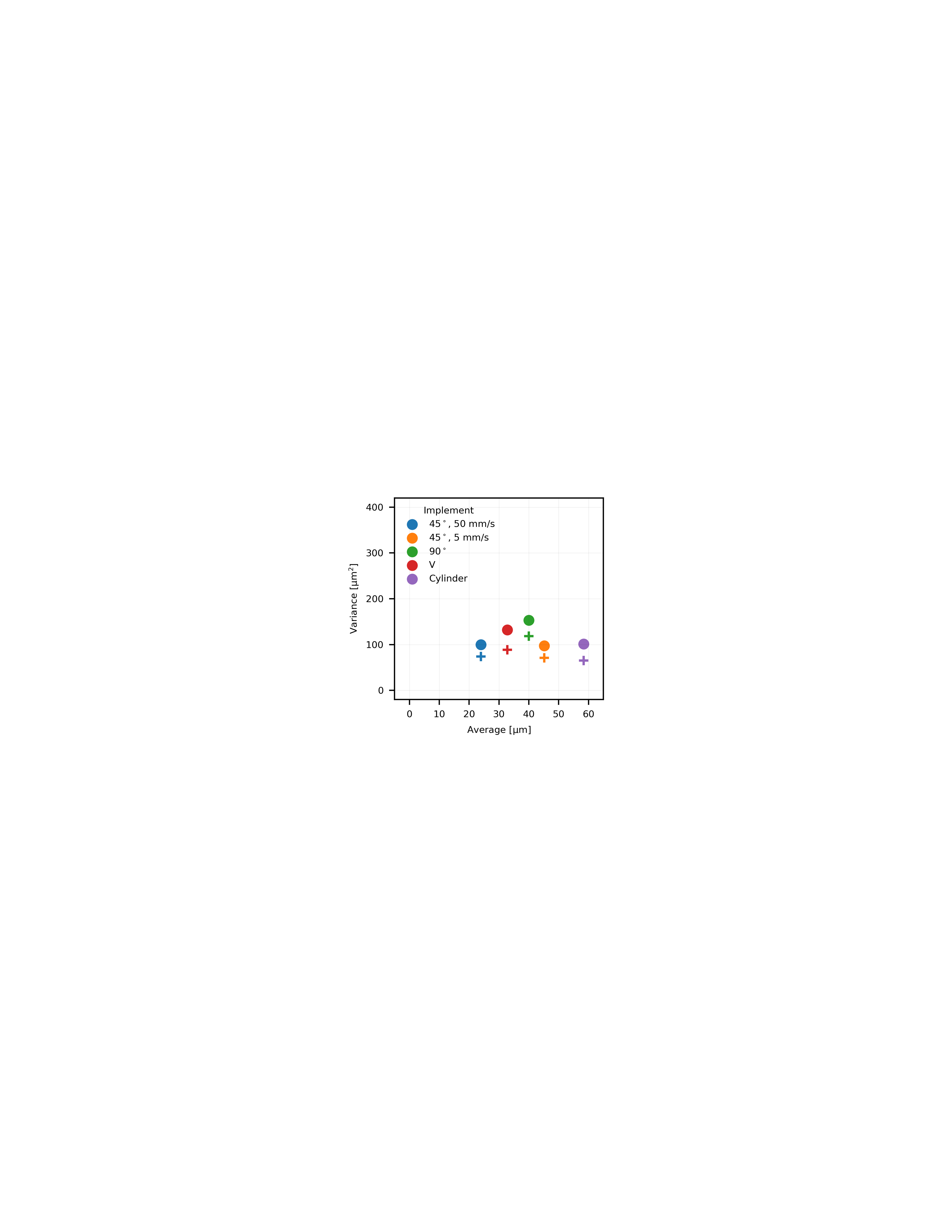}
	}
\end{center}
\vspace{-11pt}
\caption{Statistics of experimental $15-45$~\textmu m Ti-6AL-4V powder layers created with a variety of blade-like implements.  Circles represent the conventional variance of effective depth of 5 specimen layers.  Crosses indicate layer-to-layer variance, or equivalently remove spatially-repeatable variation from the prior calculation.}

\label{fig:ResultsStats}
\end{figure}

\begin{figure*}[ht]
\begin{center}
	{
	\includegraphics[trim = {1.5in 2in 1.5in 1.8in}, clip, scale=1, keepaspectratio=true]{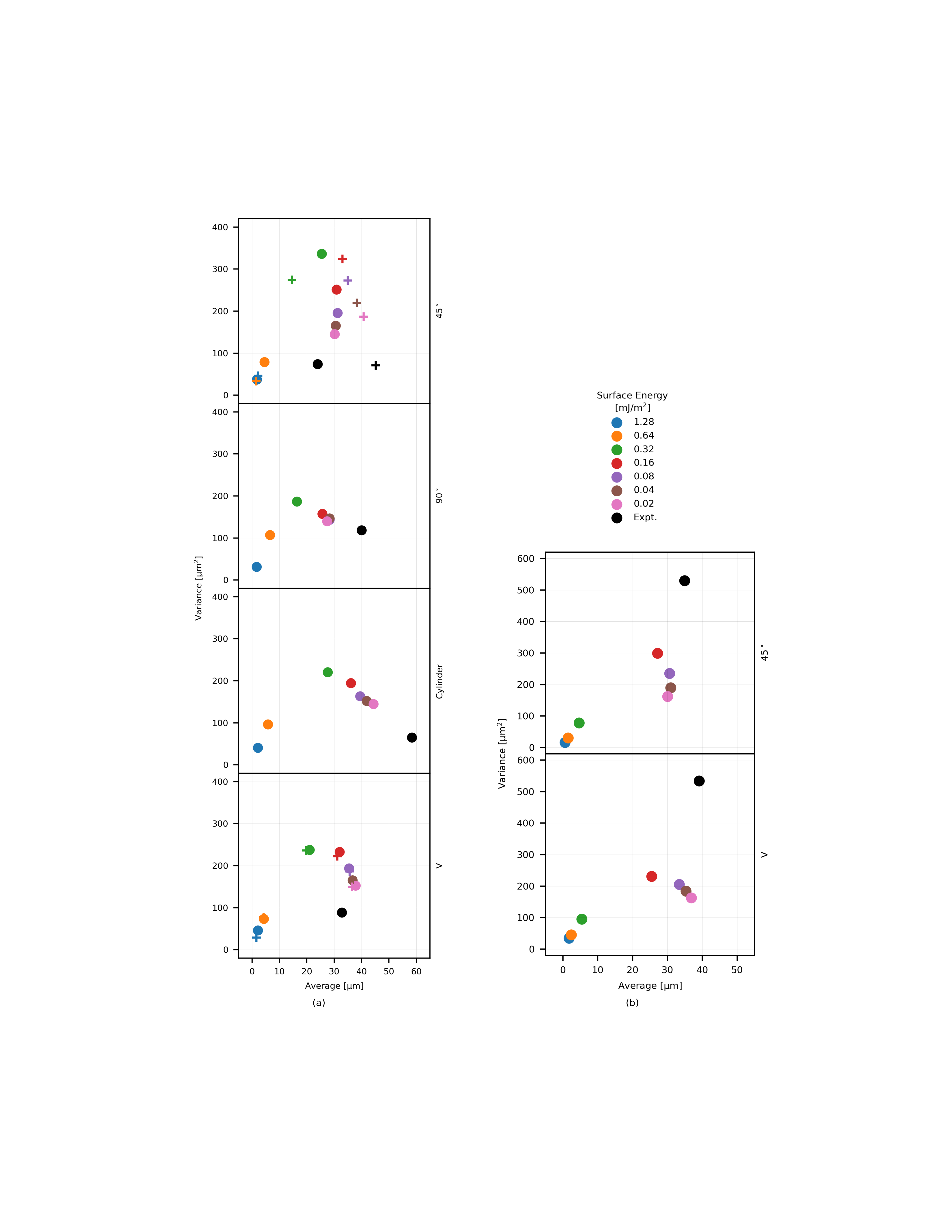}
	}
\end{center}
\vspace{-11pt}
\caption{Comparison of layer effective depth statistics for simulated powder spreading of pseudo-materials with surface energies spanning $0.02$ and $1.28$~mJ/m$^2$.  (a) Material density representative of Ti-6Al-4V.  Note: The top panel further compares layers spread at $50$ and $5$~mm/s traverse speed (circles and crosses, respectively), and the bottom panel compares V blades made from high (circles) and low compliance materials (crosses). (b) Material density representative of Al-10Si-Mg.}
\label{fig:ResultsSimsStats}
\end{figure*}

\begin{figure*}[ht]
\begin{center}
	{
	\includegraphics[trim = {1.5in 2in 1.5in 1.3in}, clip, scale=1, keepaspectratio=true]{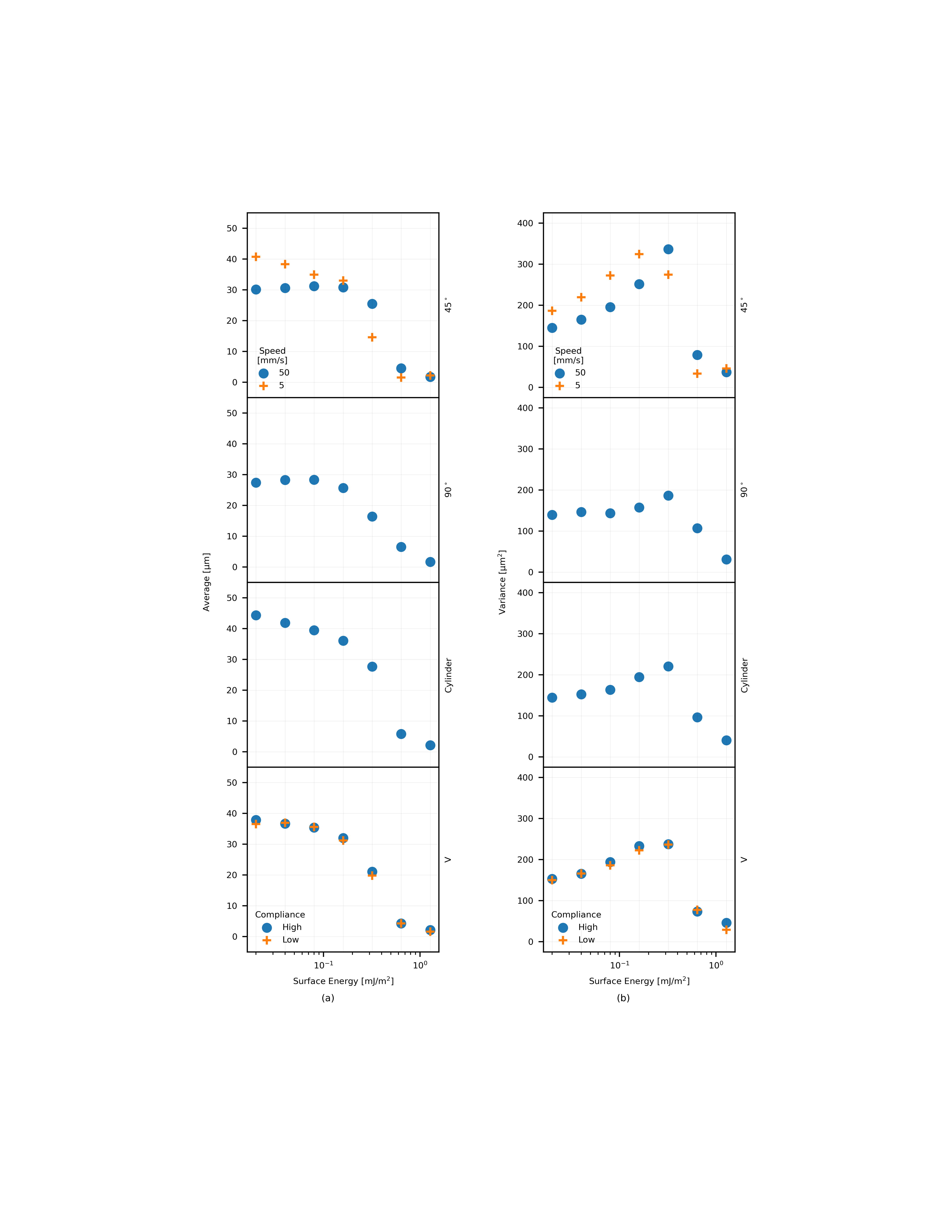}
	}
\end{center}
\vspace{-11pt}
\caption{Alternative presentation of the post-processed simulation data comprising Fig.~\ref{fig:ResultsSimsStats}a.  (a) Average effective depth.  (b) Variance of effective depth.  Note: Top panels further compare layers spread at $5$ and $50$~mm/s traverse speed, and bottom panels compare V blades made from high and low compliance materials.}
\label{fig:ResultsSimsAdhesion}
\end{figure*}

\subsubsection{Summary of Results}

Through the joint experimental and computational investigation above, the nature of powder flow and layer attributes are clearly strong functions of the powder properties and implement geometry and material.  We use layers spread with the $45^\circ$ blade as a basis for comparison, where experimental results indicate a nominally $100$~\textmu m thick powder layer features a mean effective depth of $24.0$~\textmu m and variance of $99.9$~\textmu m$^2$ when spreading $15-45$~\textmu m Ti-6Al-4V.  Simulations indicate that as surface energy of these powders is artificially increased (i.e., from $0.04$ to $0.08$~mJ/m$^2$ that is characteristic of the physical Ti-6Al-4V powder) deposition falls and variance increases up to a threshold surface energy of approx $0.32$~mJ/m$^2$, where the powder becomes too cohesive to effectively deposit.  Real world spreading performance using the $90^\circ$ blade blade is shown to be comparatively poor for practical reasons, namely that large particles induce severe streak-like defects if they become trapped under the extended flat surface of the blade.  Critically, however, we note that where deposition does occur the layers are quite uniform as evidenced by PSD analysis that shows the large surface area in contact with the powder serves to level and redistribute particles into a consistent thickness along the spreading direction.  This view is also supported by the simulation results, wherein the variance of powder layers spread with this implement and comprising pseudo materials with surface energies below the spreadability threshold than when using the $45^\circ$ blade.  Spreading with the V blade resolves the powder trapping problem exhibited by the $90^\circ$ blade blade, while still increasing the duration of powder-implement contact via radiused corner and thereby reducing layer variance.  Experimental Ti-6Al-4V layers spread with this implement show an effective depth $\approx 10$~\textmu m higher than the $45^\circ$ blade.  However, layer variance is also slightly increased, largely due to form errors of the blade edge that are transferred to the powder layer.  Counterpart simulations clarify that the increased deposition does not arise from large-scale defection of the spreading blade, but rather result from the radiused corner of the implement that serves to compress deposited powder particles into a dense configuration.  Moreover, simulations and experiments agree that this geometry is most suitable for spreading moderately cohesive powders, in that it combines the low sensitivity to surface energy of the $45^\circ$ blade and low variance from extended surface contact of the tools.  Finally, spreading with a cylinder implement is shown to result in exceptionally dense and uniform powder layers.  However, high normal forces are evidenced in the experimental powder layers by stick-slip powder flow, paralleling other studies described in the introduction, and the risk of damage to the underlying component makes such an implement impractical.

\subsection{Effect of Spreading Traverse Speed}

\begin{figure*}[ht]
\begin{center}
	{
	\includegraphics[trim = {1.7in 3in 1.7in 2.9in}, clip, scale=1, keepaspectratio=true]{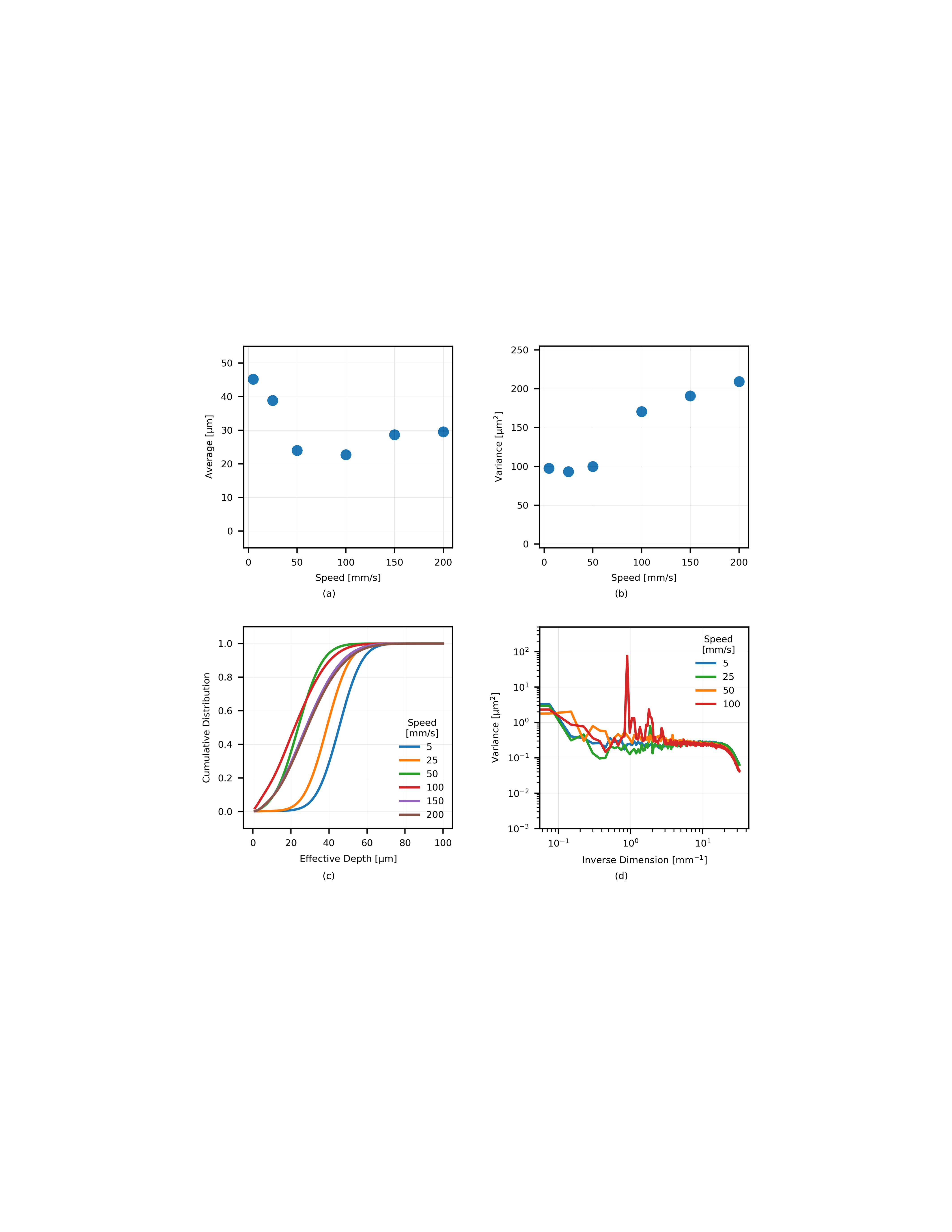}
	}
\end{center}
\vspace{-11pt}
\caption{Experimental layer statistics as a function of traverse speed.  (a) and (b) Statistics of layer effective depth.  (c) Cumulative distributions of effective depth.  (d) PSD analysis of non-uniformity size scale.}
\label{fig:ResultsTraverse}
\end{figure*}

Finally, we consider the effect of traverse speed using the combination of $45^\circ$ rigid blade and $15-45$~\textmu m Ti-6Al-4V powder.  Experimental data compare statistics from 5 layers spread at each of $5$~mm/s and $50$~mm/s (see comparisons in Figs.~\ref{fig:ResultsStats}, \ref{fig:ResultsHists_Expt}, and~\ref{fig:ResultsPSD}), in addition to single layers spread at $25$, $100$, $150$ and $200$~mm/s.  Figures~\ref{fig:ResultsTraverse}a and~\ref{fig:ResultsTraverse}b plot the average and variance of layer effective depth against traverse speed, with the full cumulative distributions shown in Fig.~\ref{fig:ResultsTraverse}c.  We find that deposition decreases as spreading speed is increased from $5$ to $50$~mm/s, and is roughly constant thereafter.  Variance in layers spread at low velocities ($5-50$~mm/s) appears largely independent of speed in Fig.~\ref{fig:ResultsTraverse}b, with a value of approximately $100$~\textmu m$^2$.  Consideration of Figs.~\ref{fig:ResultsStats} and~\ref{fig:ResultsHists_Expt} further suggests that both spatially-repeatable and layer-to-layer effective depth variation are also roughly constant for these speeds.  Increasing traverse speed to the range of $100-200$~mm/s results in a sharp increase in layer variance and modest changes in deposition.  This change in spreading behavior manifests from the coupling of machine vibrations to the powder flow, as opposed to powder kinematics \textit{per se}.  Specifically, deposition at high traverse speeds ($100$ to $200$~mm/s) is primarily observed to have alternating stripes of moderate and low effective depth, i.e., as transverse banding more severe than seen in Fig.~\ref{fig:IntroLayers}a.

Power spectral density analysis parallels these observations.  For example, Fig.~\ref{fig:ResultsPSD} shows that the extremely modest reduction in variance associated with layers spread at $5$~mm/s versus $50$~mm/s is not associated with a specific length scale, but that generally variance shows a similar functional form and is reduced at all length scales except the lowest inverse dimension bin.  Fig.~\ref{fig:ResultsTraverse}d includes data from the single $25$ and $100$~mm/s layers, where the aforementioned machine vibrations at $100$~mm/s are clearly evidenced by the sharp peaks in the effective depth PSD.  Data at higher traverse speeds are omitted here for clarity, but qualitatively resemble these results with increasingly large peaks associated with corresponding deposition patterns and machine resonances.

Simulations of spreading the $15-45$~\textmu m Ti-6Al-4V feedstock with the $45^\circ$ blade (see Figs.~\ref{fig:ResultsSimsStats}a and~\ref{fig:ResultsSimsAdhesion}) agree that spreading at a lower ($5$~mm/s) traverse speed improves layer density for powders of moderate to high flowability as compared to the $50$~mm/s baseline.  However, they also suggest that the higher particle density enables more complex arrangements thereof, and this drives the greater variance generally associated with the low-speed results.  More broadly, simulations show that deposition of powders with high surface energy ($\geq 0.32$~mJ/m$^2$) benefits from increased shear experienced by the powder pile when spread at the higher speed.

\section{Conclusion}

Transmission X-ray mapping of powder deposition provides new metrics for assessing layer quality, and several conclusions may be drawn from analysis of specimen layers created with a selection of spreading tools.  Blade geometry is shown to critically influence layer uniformity, as particles trapped under a square ($90^\circ$) blade edge, against the rigid underlying material (simulated build area), cause severe streaking artifacts.  In comparison to a $45^\circ$ knife edge, a V-shaped compliant blade is observed to modestly increase the average and variance in deposition of a nominally $15-45$~\textmu m Ti-6Al-4V powder.  Similarly, deposition and area covered increases when spreading a more cohesive powder, $20-63$~\textmu m Al-10Si-Mg, with a minimal increase in total layer variance when using this implement.  However, power spectral density (PSD) analysis of layer variance suggests that neither tool is effective in mitigating the substantial cluster-induced variance associated with this powder.  Finally, DEM simulations clarify that the primary difference in spreading between the $45^\circ$ rigid blade and the V blade is the slight bevel on the latter.  The advantage afforded by the compliant blade is in enabling a radiused geometry, improving the amount of surface area in contact between the blade and powder pile, while mitigating the high forces associated with a rigid tool of equivalent shape.  Finally, we find that, in addition to highlighting the importance of machine compliance and vibration under forces associated with high spreading speeds, average layer depth falls with increasing spreading speed, demonstrating a tradeoff between throughput and quality.

It is clearly impossible to catalog all combinations of spreading tool, powder, and boundary conditions that are of industrial relevance; nevertheless, we identify three areas of particularly impactful future work.  First, experimental and computational results, in agreement with prior research, indicate that the geometry and traverse speed of the spreading tool may be selected as to improve the stability of powder flow and the uniformity of powder layer in turn for specific feedstocks of interest.  This will critically enable the use of finer, extremely cohesive powders and smaller layer thicknesses, thereby improving the vertical resolution of powder bed AM.  Second, performance of various spreading tools in the presence of disturbances should be studied, as to clarify the role of underlying surface roughness and abrupt changes in underlying material, e.g., at the transition from spreading over powder to spreading over previously fused powder.  Finally, we envision correlating the effective depth data collected with this technique to optical image data.  If sufficiently strong correlations may be deduced, feedforward control of the AM process such as laser powder bed fusion will be possible in response to expected layer effective depth, thereby improving component quality.

\section{Acknowledgements}

Financial support was provided by: Honeywell Federal Manufacturing \& Technologies (FM\&T); a gift from Robert Bosch, LLC; a MathWorks MIT Mechanical Engineering Fellowship (to R.W.P.); and a NASA Space Technology Research Fellowship (to D.O.). We also thank Rachel Grodsky (Honeywell FM\&T) for performing laser diffraction particle sizing measurements.  P.P., C.M., and W.W. acknowledge funding of this work by the Deutsche Forschungsgemeinschaft (DFG, German Research Foundation) within project 414180263.

\beginsupplement
\FloatBarrier
\section{Layer Statistics v. Adhesion}

\section{Computational Modeling Parameters}

\begin{table*}[htb]
\centering
\caption{DEM model parameters}
\label{tab:DEMtarameters}
\renewcommand{\arraystretch}{0.7}
\begin{tabular}{ l l l } 
 \toprule
 Parameter & Value & Unit \\ 
 \midrule
 \multirow{2}{*}{Density}       &Ti-6Al-4V: 4430      &kg/m$^3$     \\ 
       &Al-10Si-Mg: 2670      &kg/m$^3$   \\ 
 Penalty parameter      &0.34      &N/m   \\ 
 Poisson's ratio     &0.342         &- \\
 Coefficient of friction     &0.4         &- \\
 Coefficient of rolling resistance     &0.07         &- \\
 Coefficient of restitution     &0.4         &- \\
 Surface energy     &varied: 0.02-1.28         &mJ/m$^2$ \\
 \midrule
 Log-normal particle size distribution:     &         & \\
 Median     &13.4968         &$\mu$m \\
 Sigma     &0.2253         &- \\
 Minimum cutoff radius     &10.1117         &$\mu$m \\
 Maximum cutoff radius     &44         &$\mu$m \\
 \bottomrule
 \end{tabular}
\end{table*}

\section*{References}

\bibliographystyle{elsarticle-num}
\bibliography{refs.bib}

\begin{thebibliography}{10}
\expandafter\ifx\csname url\endcsname\relax
  \def\url#1{\texttt{#1}}\fi
\expandafter\ifx\csname urlprefix\endcsname\relax\def\urlprefix{URL }\fi
\expandafter\ifx\csname href\endcsname\relax
  \def\href#1#2{#2} \def\path#1{#1}\fi

\bibitem{Vock2019}
S.~Vock, B.~Kl{\"o}den, A.~Kirchner, T.~Wei{\ss}g{\"a}rber, B.~Kieback, Powders
  for powder bed fusion: a review, Progress in Additive Manufacturing (2019).
\newblock \href {https://doi.org/10.1007/s40964-019-00078-6}
  {\path{doi:10.1007/s40964-019-00078-6}}.

\bibitem{Beitz2019}
S.~Beitz, R.~Uerlich, T.~Bokelmann, A.~Diener, T.~Vietor, A.~Kwade, Influence
  of powder deposition on powder bed and specimen properties, Materials 12~(2)
  (2019) 297.

\bibitem{Ziaee2019}
M.~Ziaee, N.~Crane, Binder jetting: A review of process, materials, and
  methods, Additive Manufacturing 28 (2019).
\newblock \href {https://doi.org/10.1016/j.addma.2019.05.031}
  {\path{doi:10.1016/j.addma.2019.05.031}}.

\bibitem{Jimenez2019}
E.~M. Jimenez, D.~Ding, L.~Su, A.~R. Joshi, A.~Singh, B.~Reeja-Jayan, J.~Beuth,
  Parametric analysis to quantify process input influence on the printed
  densities of binder jetted alumina ceramics, Additive Manufacturing 30 (2019)
  100864.
\newblock \href {https://doi.org/10.1016/j.addma.2019.100864}
  {\path{doi:10.1016/j.addma.2019.100864}}.

\bibitem{Korner2016}
C.~K{\"o}rner, Additive manufacturing of metallic components by selective
  electron beam melting — a review, International Materials Reviews 61~(5)
  (2016) 361--377.
\newblock \href {https://doi.org/10.1080/09506608.2016.1176289}
  {\path{doi:10.1080/09506608.2016.1176289}}.

\bibitem{Brika2020}
S.~E. Brika, M.~Letenneur, C.~A. Dion, V.~Brailovski, Influence of particle
  morphology and size distribution on the powder flowability and laser powder
  bed fusion manufacturability of ti-6al-4v alloy, Additive Manufacturing 31
  (2020) 100929.
\newblock \href {https://doi.org/https://doi.org/10.1016/j.addma.2019.100929}
  {\path{doi:https://doi.org/10.1016/j.addma.2019.100929}}.

\bibitem{Wischeropp2019}
T.~M. Wischeropp, C.~Emmelmann, M.~Brandt, A.~Pateras, Measurement of actual
  powder layer height and packing density in a single layer in selective laser
  melting, Additive Manufacturing 28 (2019) 176 -- 183.
\newblock \href {https://doi.org/https://doi.org/10.1016/j.addma.2019.04.019}
  {\path{doi:https://doi.org/10.1016/j.addma.2019.04.019}}.

\bibitem{Kiani2020}
P.~Kiani, U.~Scipioni~Bertoli, A.~D. Dupuy, K.~Ma, J.~M. Schoenung, A
  statistical analysis of powder flowability in metal additive manufacturing,
  Advanced Engineering Materials 22~(10) (2020) 2000022.
\newblock \href {https://doi.org/https://doi.org/10.1002/adem.202000022}
  {\path{doi:https://doi.org/10.1002/adem.202000022}}.

\bibitem{Avrampos2022}
P.~Avrampos, G.-C. Vosniakos,
  \href{https://www.sciencedirect.com/science/article/pii/S1526612521009002}{A
  review of powder deposition in additive manufacturing by powder bed fusion},
  Journal of Manufacturing Processes 74 (2022) 332--352.
\newblock \href {https://doi.org/https://doi.org/10.1016/j.jmapro.2021.12.021}
  {\path{doi:https://doi.org/10.1016/j.jmapro.2021.12.021}}.
\newline\urlprefix\url{https://www.sciencedirect.com/science/article/pii/S1526612521009002}

\bibitem{Mindt2016}
H.~W. Mindt, M.~Megahed, N.~P. Lavery, M.~A. Holmes, S.~G.~R. Brown, Powder bed
  layer characteristics: The overseen first-order process input, Metallurgical
  and Materials Transactions A 47~(8) (2016) 3811--3822.
\newblock \href {https://doi.org/10.1007/s11661-016-3470-2}
  {\path{doi:10.1007/s11661-016-3470-2}}.

\bibitem{Brandt2017}
M.~Brandt, The role of lasers in additive manufacturing, in: M.~Brandt (Ed.),
  Laser Additive Manufacturing, Woodhead Publishing Series in Electronic and
  Optical Materials, Woodhead Publishing, 2017, pp. 1 -- 18.
\newblock \href
  {https://doi.org/https://doi.org/10.1016/B978-0-08-100433-3.02001-7}
  {\path{doi:https://doi.org/10.1016/B978-0-08-100433-3.02001-7}}.

\bibitem{Sutton2016}
A.~T. Sutton, C.~S. Kriewall, M.~C. Leu, J.~W. Newkirk, Powders for additive
  manufacturing processes: Characterization techniques and effects on part
  properties, Solid Freeform Fabrication 1 (2016) 1004--1030.

\bibitem{Yablokova2015RheologicalImplants}
G.~Yablokova, M.~Speirs, J.~Van~Humbeeck, J.-P. Kruth, J.~Schrooten, R.~Cloots,
  F.~Boschini, G.~Lumay, J.~Luyten, {Rheological behavior of {$\beta$}-Ti and
  NiTi powders produced by atomization for SLM production of open porous
  orthopedic implants}, Powder Technology 283 (2015) 199--209.
\newblock \href {https://doi.org/10.1016/J.POWTEC.2015.05.015}
  {\path{doi:10.1016/J.POWTEC.2015.05.015}}.

\bibitem{Chen2017}
H.~Chen, Q.~Wei, S.~Wen, Z.~Li, Y.~Shi, Flow behavior of powder particles in
  layering process of selective laser melting: Numerical modeling and
  experimental verification based on discrete element method, International
  Journal of Machine Tools and Manufacture 123 (2017) 146 -- 159.
\newblock \href
  {https://doi.org/https://doi.org/10.1016/j.ijmachtools.2017.08.004}
  {\path{doi:https://doi.org/10.1016/j.ijmachtools.2017.08.004}}.

\bibitem{Meier2019CriticalManufacturing}
C.~Meier, R.~Weissbach, J.~Weinberg, W.~A. Wall, A.~J. Hart, Critical
  influences of particle size and adhesion on the powder layer uniformity in
  metal additive manufacturing, Journal of Materials Processing Technology 266
  (2019) 484 -- 501.
\newblock \href
  {https://doi.org/https://doi.org/10.1016/j.jmatprotec.2018.10.037}
  {\path{doi:https://doi.org/10.1016/j.jmatprotec.2018.10.037}}.

\bibitem{Meier2019ModelingSimulations}
C.~Meier, R.~Weissbach, J.~Weinberg, W.~A. Wall, A.~J. Hart, Modeling and
  characterization of cohesion in fine metal powders with a focus on additive
  manufacturing process simulations, Powder Technology 343 (2019) 855 -- 866.
\newblock \href {https://doi.org/https://doi.org/10.1016/j.powtec.2018.11.072}
  {\path{doi:https://doi.org/10.1016/j.powtec.2018.11.072}}.

\bibitem{Gong2014}
H.~Gong, K.~Rafi, H.~Gu, T.~Starr, B.~Stucker, Analysis of defect generation in
  ti-6al-4v parts made using powder bed fusion additive manufacturing
  processes, Additive Manufacturing 1-4 (2014) 87 -- 98.
\newblock \href {https://doi.org/https://doi.org/10.1016/j.addma.2014.08.002}
  {\path{doi:https://doi.org/10.1016/j.addma.2014.08.002}}.

\bibitem{Budding2013}
A.~Budding, T.~Vaneker, New strategies for powder compaction in powder-based
  rapid prototyping techniques, Procedia CIRP 6 (2013) 527 -- 532, proceedings
  of the Seventeenth CIRP Conference on Electro Physical and Chemical Machining
  (ISEM).
\newblock \href {https://doi.org/https://doi.org/10.1016/j.procir.2013.03.100}
  {\path{doi:https://doi.org/10.1016/j.procir.2013.03.100}}.

\bibitem{Bai2017}
Y.~Bai, G.~Wagner, C.~B. Williams, {Effect of Particle Size Distribution on
  Powder Packing and Sintering in Binder Jetting Additive Manufacturing of
  Metals}, Journal of Manufacturing Science and Engineering 139~(8) (06 2017).
\newblock \href {https://doi.org/10.1115/1.4036640}
  {\path{doi:10.1115/1.4036640}}.

\bibitem{Spierings2011}
A.~B. Spierings, N.~Herres, G.~Levy, Influence of the particle size
  distribution on surface quality and mechanical properties in am steel parts,
  Rapid Prototyping Journal 17~(3) (2011) 195--202.

\bibitem{Nan2018}
W.~Nan, M.~Pasha, T.~Bonakdar, A.~Lopez, U.~Zafar, S.~Nadimi, M.~Ghadiri,
  \href{https://www.sciencedirect.com/science/article/pii/S0032591018305278}{Jamming
  during particle spreading in additive manufacturing}, Powder Technology 338
  (2018) 253--262.
\newblock \href {https://doi.org/https://doi.org/10.1016/j.powtec.2018.07.030}
  {\path{doi:https://doi.org/10.1016/j.powtec.2018.07.030}}.
\newline\urlprefix\url{https://www.sciencedirect.com/science/article/pii/S0032591018305278}

\bibitem{Mussatto2021}
A.~Mussatto, R.~Groarke, A.~O’Neill, M.~A. Obeidi, Y.~Delaure, D.~Brabazon,
  Influences of powder morphology and spreading parameters on the powder bed
  topography uniformity in powder bed fusion metal additive manufacturing,
  Additive Manufacturing 38 (2021) 101807.
\newblock \href {https://doi.org/https://doi.org/10.1016/j.addma.2020.101807}
  {\path{doi:https://doi.org/10.1016/j.addma.2020.101807}}.

\bibitem{Kleszczynski2012}
S.~Kleszczynski, J.~Zur~Jacobsm{\"u}hlen, J.~Sehrt, G.~Witt, Error detection in
  laser beam melting systems by high resolution imaging, in: Proceedings of the
  Solid Freeform Fabrication Symposium, Vol. 2012, 2012, pp. 975--987.

\bibitem{Hendriks2019}
A.~Hendriks, R.~Ramokolo, C.~Ngobeni, M.~Moroko, D.~Naidoo, {Layer-wise powder
  deposition defect detection in additive manufacturing}, in: B.~Gu,
  H.~Helvajian, H.~Chen (Eds.), Laser 3D Manufacturing VI, Vol. 10909,
  International Society for Optics and Photonics, SPIE, 2019, pp. 99 -- 111.
\newblock \href {https://doi.org/10.1117/12.2509571}
  {\path{doi:10.1117/12.2509571}}.

\bibitem{Dana2019}
M.~Dana, I.~Zetkova, P.~Hanzl, The influence of a ceramic recoater blade on 3d
  printing using direct metal laser sintering, Manufacturing Technology Journal
  19~(1) (2019) 23--28.
\newblock \href {https://doi.org/10.21062/ujep/239.2019/a/1213-2489/MT/19/1/23}
  {\path{doi:10.21062/ujep/239.2019/a/1213-2489/MT/19/1/23}}.

\bibitem{Jacob2016}
G.~Jacob, A.~Donmez, J.~Slotwinski, S.~Moylan, Measurement of powder bed
  density in powder bed fusion additive manufacturing processes, Measurement
  Science and Technology 27~(11) (2016) 115601.
\newblock \href {https://doi.org/10.1088/0957-0233/27/11/115601}
  {\path{doi:10.1088/0957-0233/27/11/115601}}.

\bibitem{Meyer2017}
L.~Meyer, A.~Wegner, G.~Witt, Influence of the ratio between the translation
  and contra-rotating coating mechanism on different laser sintering materials
  and their packing density, in: Solid Freeform Fabrication 2017: Proceedings
  of the 28th Annual International Solid Freeform Fabrication Symposium An
  Additive Manufacturing Conference, 2017, pp. 1432--1447.

\bibitem{Chen2019}
H.~Chen, Q.~Wei, Y.~Zhang, F.~Chen, Y.~Shi, W.~Yan, Powder-spreading mechanisms
  in powder-bed-based additive manufacturing: Experiments and computational
  modeling, Acta Materialia 179 (2019) 158--171.
\newblock \href {https://doi.org/https://doi.org/10.1016/j.actamat.2019.08.030}
  {\path{doi:https://doi.org/10.1016/j.actamat.2019.08.030}}.

\bibitem{Boley2016MetalExperiment}
C.~D. Boley, S.~C. Mitchell, A.~M. Rubenchik, S.~S.~Q. Wu, {Metal powder
  absorptivity: modeling and experiment}, Applied Optics (2016).
\newblock \href {https://doi.org/10.1364/AO.55.006496}
  {\path{doi:10.1364/AO.55.006496}}.

\bibitem{Tan2017AnProcess}
J.~H. Tan, W.~L.~E. Wong, K.~W. Dalgarno, {An overview of powder granulometry
  on feedstock and part performance in the selective laser melting process},
  Additive Manufacturing 18 (2017) 228--255.
\newblock \href {https://doi.org/10.1016/J.ADDMA.2017.10.011}
  {\path{doi:10.1016/J.ADDMA.2017.10.011}}.

\bibitem{Snow2019}
Z.~Snow, R.~Martukanitz, S.~Joshi, On the development of powder spreadability
  metrics and feedstock requirements for powder bed fusion additive
  manufacturing, Additive Manufacturing 28 (2019) 78 -- 86.
\newblock \href {https://doi.org/10.1016/j.addma.2019.04.017}
  {\path{doi:10.1016/j.addma.2019.04.017}}.

\bibitem{Liu2019}
P.~Liu, Z.~Wang, Y.~Xiao, M.~F. Horstemeyer, X.~Cui, L.~Chen,
  \href{http://www.sciencedirect.com/science/article/pii/S2214860418304706}{Insight
  into the mechanisms of columnar to equiaxed grain transition during metallic
  additive manufacturing}, Additive Manufacturing 26 (2019) 22 -- 29.
\newblock \href {https://doi.org/https://doi.org/10.1016/j.addma.2018.12.019}
  {\path{doi:https://doi.org/10.1016/j.addma.2018.12.019}}.
\newline\urlprefix\url{http://www.sciencedirect.com/science/article/pii/S2214860418304706}

\bibitem{Chen2020}
H.~Chen, Y.~Chen, Y.~Liu, Q.~Wei, Y.~Shi, W.~Yan, Packing quality of powder
  layer during counter-rolling-type powder spreading process in additive
  manufacturing, International Journal of Machine Tools and Manufacture 153
  (2020) 103553.
\newblock \href
  {https://doi.org/https://doi.org/10.1016/j.ijmachtools.2020.103553}
  {\path{doi:https://doi.org/10.1016/j.ijmachtools.2020.103553}}.

\bibitem{Zhang2016}
B.~Zhang, J.~Ziegert, F.~Farahi, A.~Davies, In situ surface topography of laser
  powder bed fusion using fringe projection, Additive Manufacturing 12 (2016)
  100--107.
\newblock \href {https://doi.org/https://doi.org/10.1016/j.addma.2016.08.001}
  {\path{doi:https://doi.org/10.1016/j.addma.2016.08.001}}.

\bibitem{TanPhuc2019AManufacturing}
L.~Tan~Phuc, M.~Seita, {A high-resolution and large field-of-view scanner for
  in-line characterization of powder bed defects during additive
  manufacturing}, Materials {\&} Design 164 (2019) 107562.
\newblock \href {https://doi.org/10.1016/J.MATDES.2018.107562}
  {\path{doi:10.1016/J.MATDES.2018.107562}}.

\bibitem{Le2021}
T.~P. Le, X.~Wang, K.~P. Davidson, J.~E. Fronda, M.~Seita, Experimental
  analysis of powder layer quality as a function of feedstock and recoating
  strategies, Additive Manufacturing (2021).
\newblock \href {https://doi.org/10.1016/j.addma.2021.101890}
  {\path{doi:10.1016/j.addma.2021.101890}}.

\bibitem{Ali2018OnProcesses}
U.~Ali, Y.~Mahmoodkhani, S.~Imani~Shahabad, R.~Esmaeilizadeh, F.~Liravi,
  E.~Sheydaeian, K.~Y. Huang, E.~Marzbanrad, M.~Vlasea, E.~Toyserkani, {On the
  measurement of relative powder-bed compaction density in powder-bed additive
  manufacturing processes}, Materials {\&} Design 155 (2018) 495--501.

\bibitem{Penny2021}
R.~W. Penny, P.~M. Praegla, M.~Ochsenius, D.~Oropeza, R.~Weissbach, C.~Meier,
  W.~A. Wall, A.~J. Hart,
  \href{https://www.sciencedirect.com/science/article/pii/S2214860421003572}{Spatial
  mapping of powder layer density for metal additive manufacturing via
  transmission x-ray imaging}, Additive Manufacturing 46 (2021) 102197.
\newblock \href {https://doi.org/https://doi.org/10.1016/j.addma.2021.102197}
  {\path{doi:https://doi.org/10.1016/j.addma.2021.102197}}.
\newline\urlprefix\url{https://www.sciencedirect.com/science/article/pii/S2214860421003572}

\bibitem{Muniz2018}
J.~A. Muñiz-Lerma, A.~Nommeots-Nomm, K.~E. Waters, M.~Brochu, A comprehensive
  approach to powder feedstock characterization for powder bed fusion additive
  manufacturing: A case study on alsi7mg, Materials 11~(12) (2018).

\bibitem{Escano2018}
L.~Escano, N.~Parab, L.~Xiong, Q.~Guo, C.~Zhao, K.~Fezzaa, W.~Everhart, T.~Sun,
  L.~Chen, Revealing particle-scale powder spreading dynamics in
  powder-bed-based additive manufacturing process by high-speed x-ray imaging,
  Scientific Reports 8 (12 2018).
\newblock \href {https://doi.org/10.1038/s41598-018-33376-0}
  {\path{doi:10.1038/s41598-018-33376-0}}.

\bibitem{DuPlessis2018}
A.~du~Plessis, I.~Yadroitsev, I.~Yadroitsava, S.~G. Le~Roux, X-ray
  microcomputed tomography in additive manufacturing: A review of the current
  technology and applications, 3D Printing and Additive Manufacturing 5~(3)
  (2018) 227--247.
\newblock \href {https://doi.org/10.1089/3dp.2018.0060}
  {\path{doi:10.1089/3dp.2018.0060}}.

\bibitem{Heim2016}
K.~Heim, F.~Bernier, R.~Pelletier, L.-P. Lefebvre, High resolution pore size
  analysis in metallic powders by x-ray tomography, Case Studies in
  Nondestructive Testing and Evaluation 6 (2016) 45--52.
\newblock \href {https://doi.org/https://doi.org/10.1016/j.csndt.2016.09.002}
  {\path{doi:https://doi.org/10.1016/j.csndt.2016.09.002}}.

\bibitem{Schmidt2020}
J.~Schmidt, E.~J. Parteli, N.~Uhlmann, N.~W{\"o}rlein, K.-E. Wirth,
  T.~P{\"o}schel, W.~Peukert, Packings of micron-sized spherical particles –
  insights from bulk density determination, x-ray microtomography and discrete
  element simulations, Advanced Powder Technology 31~(6) (2020) 2293--2304.
\newblock \href {https://doi.org/https://doi.org/10.1016/j.apt.2020.03.018}
  {\path{doi:https://doi.org/10.1016/j.apt.2020.03.018}}.

\bibitem{Oropeza2022}
D.~Oropeza, R.~W. Penny, D.~Gilbert, A.~J. Hart,
  \href{https://www.sciencedirect.com/science/article/pii/S0032591021010718}{Mechanized
  spreading of ceramic powder layers for additive manufacturing characterized
  by transmission x-ray imaging: Influence of powder feedstock and spreading
  parameters on powder layer density}, Powder Technology 398 (2022) 117053.
\newblock \href {https://doi.org/https://doi.org/10.1016/j.powtec.2021.117053}
  {\path{doi:https://doi.org/10.1016/j.powtec.2021.117053}}.
\newline\urlprefix\url{https://www.sciencedirect.com/science/article/pii/S0032591021010718}

\bibitem{Haeri2017}
S.~Haeri, Y.~Wang, O.~Ghita, J.~Sun, Discrete element simulation and
  experimental study of powder spreading process in additive manufacturing,
  Powder Technology 306 (2017) 45 -- 54.
\newblock \href {https://doi.org/10.1016/j.powtec.2016.11.002}
  {\path{doi:10.1016/j.powtec.2016.11.002}}.

\bibitem{Haeri2017_2}
S.~Haeri, Optimisation of blade type spreaders for powder bed preparation in
  additive manufacturing using dem simulations, Powder Technology 321 (2017) 94
  -- 104.
\newblock \href {https://doi.org/10.1016/j.powtec.2017.08.011}
  {\path{doi:10.1016/j.powtec.2017.08.011}}.

\bibitem{Parteli2016}
E.~J. Parteli, T.~P{\"o}schel, Particle-based simulation of powder application
  in additive manufacturing, Powder Technology 288 (2016) 96--102.
\newblock \href {https://doi.org/https://doi.org/10.1016/j.powtec.2015.10.035}
  {\path{doi:https://doi.org/10.1016/j.powtec.2015.10.035}}.

\bibitem{Wang2021}
L.~Wang, A.~Yu, E.~Li, H.~Shen, Z.~Zhou, Effects of spreader geometry on powder
  spreading process in powder bed additive manufacturing, Powder Technology 384
  (2021) 211--222.
\newblock \href {https://doi.org/https://doi.org/10.1016/j.powtec.2021.02.022}
  {\path{doi:https://doi.org/10.1016/j.powtec.2021.02.022}}.

\bibitem{Wang2020}
L.~Wang, E.~Li, H.~Shen, R.~Zou, A.~Yu, Z.~Zhou,
  \href{https://www.sciencedirect.com/science/article/pii/S0032591019311532}{Adhesion
  effects on spreading of metal powders in selective laser melting}, Powder
  Technology 363 (2020) 602--610.
\newblock \href {https://doi.org/https://doi.org/10.1016/j.powtec.2019.12.048}
  {\path{doi:https://doi.org/10.1016/j.powtec.2019.12.048}}.
\newline\urlprefix\url{https://www.sciencedirect.com/science/article/pii/S0032591019311532}

\bibitem{Oropeza2021}
D.~Oropeza, R.~Roberts, A.~J. Hart, A modular testbed for mechanized spreading
  of powder layers for additive manufacturing, Review of Scientific Instruments
  92~(1) (2021) 015114.
\newblock \href {https://doi.org/10.1063/5.0031191}
  {\path{doi:10.1063/5.0031191}}.

\bibitem{Gent1958}
A.~N. Gent, On the relation between indentation hardness and young's modulus,
  Rubber Chemistry and Technology 31~(4) (1958) 896--906.

\bibitem{ASTM2013_ReposeAngle}
ASTM, Standard test methods for flow rate of metal powders using the hall
  flowmeter funnel, Tech. rep., ASTM International West Conshohocken, PA
  (2013).

\bibitem{Hausner1967}
H.~H. Hausner, Friction conditions in a mass of metal powder., International
  Journal of Powder Metallurgy (1 1967).

\bibitem{Birch1979}
R.~Birch, M.~Marshall, Computation of bremsstrahlung x-ray spectra and
  comparison with spectra measured with a ge(li) detector, Physics in Medicine
  and Biology 24~(3) (1979) 505--517.
\newblock \href {https://doi.org/10.1088/0031-9155/24/3/002}
  {\path{doi:10.1088/0031-9155/24/3/002}}.

\bibitem{Lambert1760}
J.-H. Lambert, Photometria, sive de Mensura et gradibus luminis, colorum et
  umbrae, sumptibus viduae E. Klett, 1760.

\bibitem{Saloman1988X-ray92}
E.~Saloman, J.~Hubbell, J.~Scofield, {X-ray attenuation cross sections for
  energies 100 eV to 100 keV and elements Z = 1 to Z = 92}, Atomic Data and
  Nuclear Data Tables 38~(1) (1988) 1--196.
\newblock \href {https://doi.org/10.1016/0092-640X(88)90044-7}
  {\path{doi:10.1016/0092-640X(88)90044-7}}.

\bibitem{Dyson1990}
N.~Dyson, X-rays in Atomic and Nuclear Physics, Cambridge University Press,
  1990.

\bibitem{Holl1988}
I.~{Holl}, E.~{Lorenz}, G.~{Mageras}, A measurement of the light yield of
  common inorganic scintillators, IEEE Transactions on Nuclear Science 35~(1)
  (1988) 105--109.
\newblock \href {https://doi.org/10.1109/23.12684}
  {\path{doi:10.1109/23.12684}}.

\bibitem{Meier2021GAMM}
C.~Meier, S.~L. Fuchs, N.~Much, J.~Nitzler, R.~W. Penny, P.~M. Praegla, S.~D.
  Pröll, Y.~Sun, R.~Weissbach, M.~Schreter, N.~E. Hodge, A.~J. Hart, W.~A.
  Wall, {Physics-Based Modeling and Predictive Simulation of Powder Bed Fusion
  Additive Manufacturing Across Length Scales}, GAMM-Mitteilungen (2021).
\newblock \href {https://doi.org/https://doi.org/10.1002/gamm.202100014}
  {\path{doi:https://doi.org/10.1002/gamm.202100014}}.

\bibitem{Baci}
Baci: A comprehensive multi-physics simulation framework,
  \url{https://baci.pages.gitlab.lrz.de/website}, accessed: 2021-06-06 (2021).

\bibitem{Derjaguin1975}
B.~Derjaguin, V.~Muller, Y.~Toporov, Effect of contact deformations on the
  adhesion of particles, Journal of Colloid and Interface Science 53~(2) (1975)
  314--326.
\newblock \href {https://doi.org/https://doi.org/10.1016/0021-9797(75)90018-1}
  {\path{doi:https://doi.org/10.1016/0021-9797(75)90018-1}}.

\bibitem{Cain2001}
R.~G. Cain, N.~W. Page, S.~Biggs, Microscopic and macroscopic aspects of
  stick-slip motion in granular shear, Physical Review E 64~(1) (2001) 016413.

\end{thebibliography}

\end{document}